\documentclass[a4paper,onecolumn]{quantumarticle}
\pdfoutput=1
\usepackage{dsfont}
\usepackage{mathtools}
\usepackage[pdftex]{graphicx}
\usepackage{epsfig}
\usepackage{dcolumn}
\usepackage{bm}
\usepackage{xcolor}
\usepackage{bbm}
\usepackage{multirow}
\usepackage[ruled,vlined]{algorithm2e}
\usepackage[english]{babel}
\newenvironment{normalize}{\leftskip-\leftmargin}{\par}

\usepackage[amsmath,thmmarks]{ntheorem}
\usepackage{amssymb}
\usepackage{amsbsy}
\usepackage[normalem]{ulem}
\theoremseparator{.}
\newtheorem{prop}{Proposition}

\newtheorem{lemma}{Lemma}

\newtheorem{remark}{Remark}

\theoremsymbol{\ensuremath{\color{lightgray}\blacksquare}}
\theorembodyfont{\upshape}
\newtheorem*{proof}{Proof}

\usepackage{url}
\usepackage[colorlinks = true,
            urlcolor=blue,
            raiselinks=true
            ]{hyperref}
\hypersetup{extension = }

\usepackage{cleveref}
\crefname{equation}{Equation}{Equations}
\crefname{proof}{Proof}{Proofs}
\creflabelformat{proof}{#2proof#3}
\crefname{remark}{Remark}{Remarks}
\crefname{prop}{Proposition}{Propositions}
\usepackage{longtable}

\usepackage[numbers,sort&compress]{natbib}

\newcommand{\ket}[1]{\left\vert#1\right\rangle}
\newcommand{\bra}[1]{\left\langle #1 \right\vert}

\newcommand{\Fd}{\mathds{F}_d}

\DeclareMathOperator{\tr}{tr} 

\renewcommand{\vec}[1]{\ensuremath{\mathbf{#1}}}

\newcommand{\id}{\mathds{1}}

\newcommand{\bl}[1]{{\color{black} #1}}

\begin{document}

\title{Finite-Function-Encoding Quantum States}

\author{Paul~Appel}
    \email{paul.appel@gmail.com}
\affiliation{Atominstitut, Technische Universit\"at Wien, Stadionallee 2, 1020 Vienna, Austria}
 \affiliation{Institute  for  Quantum  Optics  and  Quantum  Information -- IQOQI  Vienna,
Austrian  Academy  of  Sciences,  Boltzmanngasse  3,  1090  Vienna,  Austria}

\author{Alexander~J.~Heilman}
\affiliation{Mathematics and Physics, Lebanon Valley College,  101 North College Avenue, Annville, Pennsylvania, 17003, United States of America}

\author{Ezekiel~W.~Wertz}
\affiliation{Mathematics and Physics, Lebanon Valley College, 101 North College Avenue, Annville, Pennsylvania, 17003, United States of America}

\author{David~W.~Lyons}
 \email{lyons@lvc.edu}
\affiliation{Mathematics and Physics, Lebanon Valley College,  101 North College Avenue, Annville, Pennsylvania, 17003, United States of America}

\author{Marcus~Huber}
    \email{marcus.huber@tuwien.ac.at}
    \affiliation{Atominstitut, Technische Universit\"at Wien, Stadionallee 2, 1020 Vienna, Austria}
 \affiliation{Institute  for  Quantum  Optics  and  Quantum  Information -- IQOQI  Vienna,
Austrian  Academy  of  Sciences,  Boltzmanngasse  3,  1090  Vienna,  Austria}

\author{Matej~Pivoluska}
 \email{pivoluskamatej@gmail.com}
\affiliation{
 Institute of Computer Science, Masaryk University, Botanick\'{a} 68a, 60200 Brno, Czech Republic
}%

\affiliation{
Institute of Physics, Slovak Academy of Sciences,
D\'{u}bravsk\'{a} cesta 5807/9, 
845 11 Karlova Ves, Slovakia
}

\author{Giuseppe~Vitagliano}
    \email{giuseppe.vitagliano@gmail.com}
    \affiliation{Atominstitut, Technische Universit\"at Wien, Stadionallee 2, 1020 Vienna, Austria}
 \affiliation{Institute  for  Quantum  Optics  and  Quantum  Information -- IQOQI  Vienna,
Austrian  Academy  of  Sciences,  Boltzmanngasse  3,  1090  Vienna,  Austria}


\begin{abstract}
\noindent
We introduce finite-function-encoding (FFE) states which
encode arbitrary $d$-valued logic functions, i.e., multivariate functions over the ring of integers modulo $d$, and investigate some of their structural properties.
We also point out some differences between polynomial and non-polynomial function encoding states:
The former can be associated to graphical objects, that we dub tensor-edge hypergraphs (TEH), which are a generalization
of hypergraphs with a tensor attached to each hyperedge encoding the coefficients of the different monomials.
To complete the framework, we also introduce a notion of finite-function-encoding Pauli (FP) operators, which correspond to elements of what is known as the generalized symmetric group in mathematics. 
First, using this machinery, we study the stabilizer group associated to FFE states and observe how qu$d$it hypergraph states introduced in Ref.~\cite{2017PhRvA..95e2340S} admit stabilizers of a particularly simpler form.  
Afterwards, we investigate the classification of FFE states under local unitaries (LU), and, after showing the complexity of this problem, we focus on the case of bipartite states and especially on the classification under local FP operations (LFP).
We find all LU and LFP classes for two qutrits and two ququarts and study several other special classes, pointing out the relation between maximally entangled FFE states and complex Butson-type Hadamard matrices.
Our investigation showcases also the relation between the properties of FFE states, especially their LU classification, and the theory of finite rings over the integers.

\end{abstract}

\maketitle

\section{Introduction}
\noindent
Higher-dimensional quantum systems  have become a common research interest in many fields of quantum information theory.
It has been shown that they exhibit potential advantages due to higher security of cryptography protocols \cite{Bruss2002,Cerf2002,HuberPawlowski2013,doi:10.1002/qute.201900038}, better key rates in QKD  \cite{islam2017provably,2020arXiv200412824D,2020arXiv201103005H}, reduced circuit complexity \cite{Wang2020}, improved quantum error correction \cite{Watson2015} and magic state distillation \cite{Campbell2012,Campbell2014}.
At the same time, the complexity of  the geometry of high-dimensional quantum states makes a \emph{general} treatment of states \bl{rapidly} unfeasible. 
This warrants the search for subsets of states which are sufficiently complex to inherit the advantages of higher dimensional quantum systems, while at the same time are easily described.
Identifying these sets is motivated by the fact that for many tasks only some states are useful, while most states are not \cite{Gross_2009}. 
\bl{Hypergraph states are}  a prominent example of such \bl{a useful}{} set of states.
 Graph states and stabilizer states have become a central  pillar of contemporary research into quantum information processing:
 appearing as underlying resources in measurement based quantum
 computation \cite{raussendorf2003measurement}; being of central
 importance to quantum algorithms \cite{2011PhRvA..83e2313B}; and providing general insight into many body entanglement \cite{PhysRevA.69.062311} and special instances of the marginal problem ~\cite{Raissi20}. 
 Hypergraph states \cite{1367-2630-15-11-113022} have received a lot of attention recently, from analytic Bell violations \cite{2016PhRvL.116g0401G} to providing more general resources for quantum computation \cite{2016npjQI...216036M,2018arXiv180512093G,2018arXiv180907552T}.
Qu\emph{b}it hypergraph states  can be seen as an encoding \bl{of} Boolean functions into the relative phases of many-qubit quantum states \cite{1367-2630-15-11-113022,2013PhRvA..87b2311Q,2018arXiv181100308D}.

In this work, we focus on this encoding and provide a framework to encode higher valued logical functions in higher dimensional quantum states.
For simplicity, we will call these functions \emph{finite functions} and the states \emph{finite function encoding (FFE) states,} respectively.
This approach turns out to be more general than trying to generalize graphs to higher dimensions, i.e., only a subset of functions, namely polynomial functions, can be associated to a graph, i.e.,  a pair of a set of vertices and a set of edges.
Note, that we use ``graph'' to mean the general object and \emph{not} simple graphs, which is often done in physics.

Let us now discuss two potential applications of the framework: the investigation of higher dimensional algorithms and the potential to compress quantum circuits.
The encoding of Boolean functions in quantum states is one of the main ingredients in ground-breaking qubit quantum algorithms by Deutsch and Jozsa \cite{doi:10.1098/rspa.1992.0167} and Grover \cite{10.1145/237814.237866}.
Our idea is that encoding higher dimensional finite functions into larger a Hilbert space could be harnessed for the improved efficiency of quantum algorithms, just as the intrinsic dimensionality of photon entanglement can be harnessed for improved communication \cite{2017PhRvL.118k0501M, Ecker_2019,2020arXiv201103005H,2020arXiv200412824D}.
Furthermore, higher-dimensional logic might prove to have an additional advantage in actual implementations of quantum algorithms thanks to its intrinsically higher expressivity---using higher dimensional Hilbert spaces allows us to encode multiple logical qubits used in a quantum circuit into a single physical system, thus reducing the number of non-local gates a quantum computer has to perform.
This reduction in the number of non-local operations is highly desirable, since they still pose a fundamental challenge in practical implementations of quantum computation: In the quantum case each two-qubit gate represents an entangling operation, which is still the fundamental challenge in practical quantum computation and therefore cannot be considered a negligibly cheap resource.
Using higher dimensional logic, many of the two-qubit gates could in principle be replaced by local gates acting on higher-dimensional quantum systems, which has a potential to greatly simplify the actual implementations using the practical quantum processors of the NISQ era.
It is for this reason that first investigations have been started into higher-dimensional quantum computation \cite{2009NatPh...5..134L,Heyfron2019}.

While there are many more potential applications, we are first interested in providing the underlying framework itself.
To that end, we define the finite-function-encoding (FFE) states in \cref{sect:definition}  and introduce a group of finite-function-encoding Pauli (FP) operators (\cref{sec:FFE_Pauli}), which in turn are used to develop a stabilizer formalism for the FFE states (\cref{sect:stab}).
The structure of these stabilizers is unsurprisingly more complicated than the one for qudit hypergraph states introduced in Ref.\cite{2017PhRvA..95e2340S}:
Only a subset of FFE states, namely those with stabilizers that can be decomposed as products of operators that commute, are related to the qudit hypergraph states.
In \cref{sec:TEH}, we discuss the intricacies of associating graphs to functions.
First, from the fact that in non-prime dimensions not every finite function is a polynomial function, we observe that \emph{only} polynomial functions can be associated to graphs.
Next, we discuss how polynomials are encoded into graphs, and we define a new graph, the tensor-edge hypergraph (TEH), to encode arbitrary polynomials.
Finally, we discuss, how these TEHs are a generalization of the previously defined qudit hypergraphs \cite{2017PhRvA..95e2340S}.
In \cref{sect:luequiv} we investigate the equivalence of FFE states under local operations, namely the previously defined finite-function-encoding Pauli operators and more general unitary operations.
First, we give a bound on the number of equivalence classes under local finite-function-encoding Pauli (LFP) operations, \bl{which shows that the}
number of classes becomes rapidly unfeasible \bl{to compute} both with \bl{increasing} local dimension and number of parties involved.
We then focus on the bipartite scenario and give a full LFP and local unitary (LU) classification for dimensions $d=3,4$ and find partial results for dimension $d=6$. We observe that bipartite maximally entangled FFE states are closely related to complex Hadamard matrices of Butson type \cite{TadejZyczkowski2006}. Using the theory of Hadamard matrices, we are able to identify several entanglement classes of FFE states and make a partial classification of states with low Schmidt rank, focusing on states which are maximally entangled in lower-dimensional subspaces.

\section{Definition of Finite-Function-Encoding quantum states}
\noindent
\label{sect:definition}
Motivation for this work begins with the observation that $n$-qubit quantum graph and hypergraph states naturally encode Boolean functions \cite{1367-2630-15-11-113022,2013PhRvA..87b2311Q,2018arXiv181100308D}. 
This motivates a generalization to encode multi-valued logical functions, which we call finite functions for brevity, in the phase of multi-qudit states of arbitrary local Hilbert space dimension $d$.
At first glance, this generalization looks straightforward: A uni-variate Boolean function $f_2$ takes inputs from the set $\{0,1\}$ and maps to itself, i.e.,  $f_2\colon\{0,1\}\to \{0,1\}$, while a uni-variate multi-valued logical function $f_d$ is the map  $f_d\colon\{0,\dots, d-1\}\to \{0,\dots, d-1\}$.
However, the underlying structure of these sets is different: Whenever $d$ is prime, the set $\{0,\dots, d-1\}$ together with addition and multiplication modulo $d$ is a \emph{finite field} $\mathds{F}_d$, when $d$ is non-prime it is the \emph{ring of integers modulo $d$}, denoted $\mathds{Z}_d$. 
The main difference to finite fields is that in rings \emph{not every} element has a multiplicative inverse, which has some profound implications, i.a., for the possibility to express a function $f_d$ as a polynomial.
Before we discuss the polynomiality of functions $f_d$ further, let us first understand how they are encoded into quantum states.
An intuitive way to think about finite functions $f_d$ is by representing every function uniquely by the tuple of its image, i.e., $f_d \leftrightarrow (f(0),f(1),\dots ,f(d-1))$.
This tuple is then encoded in a quantum  state $\ket{f_d}$ by associating the phase of a computational basis element labeled $\ket{i}$ with $f_d(i)$; $\ket{f_d}= (\omega_d)^{f(0)}\ket{0}+(\omega_d)^{f(1)}\ket{1}+\dots + \omega_d^{f(d-1)}\ket{d-1}$, where $\omega_d=e^{2\pi i/d}$ is the $d$-th principal complex root of unity.
However, encoding  uni-variate functions corresponds only to local quantum states.
More interesting are the $n$-partite quantum states with local dimension $d$, which encode $n$-variate $d$-valued finite functions $f_d: \mathds{Z}_d^{n} \to \mathds{Z}_d$.
Again, we can identify the function $f_d$ with its image as $f_d\leftrightarrow(f_d(\vec{e}))_{\forall \vec{e}\in\mathds{Z}_d^{ n}}$ and proceed to identify the image with a pure quantum state.
\begin{equation}
  \label{dooleanfnencstate}
  \ket{f_d}=\frac{1}{\sqrt{d^n}}\sum_{\vec{x}\in \mathds Z_d^n}
\omega_d^{f_d(\vec{x})} \ket{\vec{x}},
\end{equation}
where $\omega_d=e^{2\pi i/d}$ is again  the $d$-th principal complex root of unity.
To simplify the notation, we will drop the subscript of  $\omega_d$ and $f_d$ whenever the local dimension is clear from the context.  
We call these states {\em finite-function-encoding} (FFE) states. 

\subsection{Finite-Function-Encoding Pauli Operations}
\label{sec:FFE_Pauli}
\noindent
Let us now define two sets of operators which together we call {\it finite-function-encoding} Pauli \bl{(FP)} operators:
Given a function $h\colon (\mathds{Z}_d)^n\bl{\to} \mathds{Z}_d$, we will write $Z_h$ to denote the diagonal operator 
\begin{equation}
    \label{genZopdef}
    Z_h = \sum_{\vec{x}\in (\mathds{\bl{Z}}_d)^n} \omega_d^{h(\vec{x})} \ket{\vec{x}}\bra{\vec{x}}.
\end{equation}
The ordinary 1-qubit Pauli $Z$ operator is the special case for $h\colon \mathds{F}_2\bl{\to}\mathds{F}_2$ given by $h(0)=0, h(1)=1$. 
\bl{Note, that the sum of two functions is represented by the product of respective function-encoding $Z$ operators}, i.e., $Z_{f+g}=Z_f Z_g$. 
Similarly, by applying a $Z_f$ gate to a FFE state $\ket{g}$ one obtains $$Z_f\ket{g}=\ket{f+g}.$$

Now let us define the {\em finite-function-encoding Pauli $X$} operations.
In this case, since we want a unitary operation, we have to consider only the permutations and associate a $X$-type operator to each of them.
Given a permutation $\pi\in \mbox{\rm Sym}(\mathds{Z}^n_d)$, where $\mbox{\rm Sym}(\mathds{Z}^n_d)$ is the symmetric group over $\mathds{Z}^n_d$, we will write $X_{\pi}$ for the operator   
\begin{equation}
    X_\pi = \sum_{\vec{x}\in \mathds{Z}^n_d} \ket{\pi(\vec{x})} \bra{\vec{x}},
\end{equation}
The ordinary $d$-dimensional Pauli $X$ operator is obtained in the case $\pi=\kappa^{+}$, where  
\begin{align} \label{plus1cycle}
\kappa^{+} &\colon k\,\bl{\mapsto} k+1 \pmod{d},
\end{align}
is a $d$-cycle. 
We will write $\pi_i\colon \mathds{Z}_d^n \bl{\to \mathds{Z}_d^n}$ to denote the permutation $\pi$ acting on the $i$th variable, i.e.,
\bl{$\pi_i(\vec{x)}
:= (x_1,x_2,\ldots,\pi(x_i),\ldots,x_n)
$} and we write $X_{\pi_i}$ to denote the corresponding (local) operator acting on the $i$-th qudit of a $n$-qudit state. 

The action of a finite-function-encoding Pauli $X_\pi$ operator on the FFE state $\ket{f}$ is  
\begin{align}\label{eq:XAction}
    X_{\pi}\ket{f} =\frac{1}{\sqrt{d^n}} \sum_\vec{x} \omega_d^{f(\vec{x})} \ket{\pi(\vec{x})} = \frac{1}{\sqrt{d^n}}\sum_{\pi^{-1}(\vec{y})} \omega_d^{f(\pi^{-1}(\vec{y}))}\ket{\vec{y}} =\frac{1}{\sqrt{d^n}} \sum_{\vec{y}} \omega_d^{f(\pi^{-1}(\vec{y}))}\ket{\vec{y}} ,  
\end{align}
where the sum over $\vec x:= \pi^{-1}(\vec{y})$ is the same as the sum over $\vec y$,  since $\pi$ is a permutation.

We call the group generated by $X_{\pi}$ and $Z_h$ operators the {\em finite-function-encoding Pauli (FP) group}; in mathematical literature, this group is called the \emph{generalized symmetric group.} 
The FP group is not Abelian: the multiplication rules and commutation relations between these operators are given by 
\begin{subequations}\label{XZalgebra}
\begin{align}
Z_f Z_g &= Z_g Z_f = Z_{f+g} \\ 
X_{\pi}X_{\sigma} &= X_{\pi\circ \sigma} \\ 
X_{\pi}Z_{h\circ \pi} &= Z_h X_{\pi} \label{XZ_commutation}  
\end{align}
\end{subequations}
At the single-particle level, we call the corresponding group the {\em Local finite-function-encoding Pauli (LFP) group}, i.e., the group generated by $X_{\pi}$ and $Z_h$ for single variable functions $h$. 
The LFP group is also a non-commutative group with similar commutation relations as Eq.~(\ref{XZalgebra}) and is a generalization of the \bl{Heisenberg-Weyl} group of single qudits.

\subsection{Stabilizers of FFE States}
\label{sect:stab}
\noindent 
In this section, we construct stabilizers for FFE states. 
In particular, we make use of finite-function-encoding Pauli operations to generalize the Pauli stabilizer formalism for hypergraph states \cite{1367-2630-15-11-113022,2017PhRvA..95e2340S}.
We present a construction of a stabilizer set that uniquely determines an arbitrary FFE state, and we show that a property
called \emph{internal commutativity} is satisfied if and only if the FFE state can be obtained with local finite-function-encoding Pauli operations from the qudit hypergraph states defined in Ref.~\cite{2017PhRvA..95e2340S}.

Recall that graph states are determined by a discrete Abelian group of local Pauli operators, generated by $X_i\bigotimes_{j\neq i} Z_j$, where  $X_i$ and $Z_j$ are local operators  acting on the $i$-th and $j$-th subsystem. 
Hypergraph states are also determined by an Abelian discrete group of stabilizers of a similar form, but this time, with controlled-$Z$ operators, which are no longer local \cite{1367-2630-15-11-113022,2017PhRvA..95e2340S}, but still commute with the $X$ operators. 
Here, we show that FFE states are also determined by Abelian groups of operators having a similar form, but in general the finite-function-encoding Pauli $Z$ operators will not commute with the $X$ operators. 
The latter {\it internal commutativity}, as we are going to prove, is satisfied only for a special class of FFE states, that are LFP equivalent to the qudit hypergraph states. 

Given a function $f\colon \mathds{Z}_d^n \bl{\to} \mathds{Z}_d$ and a $n$-$d$it permutation $\pi\colon \mathds{Z}^n_d\bl{\to} \mathds{Z}^n_d$, let $S_{f,\pi}$ denote the operator 
\begin{equation}\label{def:stabilizer}
S_{f,\pi}=X_{\pi}Z_{f\circ \pi -f}.
\end{equation}
It is straightforward to check that $S_{f,\pi}$ stabilizes $\ket{f}$, that is
\[
S_{f,\pi} \ket{f} = \ket{f}.
\]
For a fixed $f$, these $(d^n)!$ operators $S_{f,\pi}$ (over all $(d^n)!$ permutations $\pi$ of $n$-$d$its) constitute the \bl{FP} stabilizer group of $\ket{f}$. This group is not Abelian, but it has some useful Abelian subgroups. 
In particular, we will show that there is an Abelian subgroup of \bl{FP} stabilizers with the property that the FFE state from which they are constructed is the unique simultaneous $+1$-eigenstate of the set of $n$ generators of the subgroup.
It is easy to see that $S_{f,\pi}$ and $S_{f,\sigma}$ commute whenever $\pi$ and $\sigma$ commute. 
In particular, $S_{f,\pi_i},S_{f,\sigma_j}$, where the permutations act on different qudits always commute.
If we fix \bl{an arbitrary} $d$-cycle ${\kappa_i}$ for each \bl{subsystem} $i$, the set of operators $\{ S_{f,\kappa_i}\}_{i=1}^n$ uniquely determines $\ket{f}$, among all $n$-qudit states, as their simultaneous $+1$-eigenstate. 

\begin{prop}\label{cyclestabunique}
The state $\ket{f}$ is the unique simultaneous $+1$-eigenvector of the set of $n$ FP operators $\{ S_{f,\kappa_i}\colon 1\leq i\leq n\}$, where for each $i$, $\kappa_i$ is a $d$-cycle.
\end{prop}
The proof can be found in Appendix~\ref{appendixA}.
In the above construction, the choice of the $d$-cycles $\kappa_i$ is left free.

Now, we are interested in another property of the stabilizers: \emph{internal commutativity}.
In other words, we are interested to find a set of stabilizing operators that determine a state $\ket{f}$ uniquely and also \emph{internally 
commute}, i.e., $
X_{\kappa_i}Z_{f\circ\kappa_{i} - f}
=Z_{f\circ\kappa_{i} - f}X_{\kappa_i}
\;\forall i$. 
This will be done by looking at the previous construction and studying which functions give rise to internally commuting stabilizing operators:

\begin{lemma}
\label{lem:commutativity_cyclic_permutation}
Let $\kappa$ be a $d$-cycle. Then $X_{\kappa_i}Z_h = Z_hX_{\kappa_i}$ if and only if $h(\vec{x})$ does not depend on the value of $x_i$.

\end{lemma}
\begin{proof}
From the general commutativity relation \eqref{XZ_commutation} we have,
\begin{align}\label{eq:commutativityofhandhkappa}
   X_{\kappa_i}Z_h &= Z_h X_{\kappa_i}\Longleftrightarrow h\circ\kappa_i(\vec{x}) = h(\vec{x}).
\end{align}
This implies  
\[
\forall c\in\Fd: h(x_1,\dots,x_{i-1},c,x_{i+1},\dots,x_n) = h(x_1,\dots,x_{i-1},\kappa_i(c),x_{i+1},\dots,x_n),
\]
which, together with  $\kappa_i$ being a $d$-cycle, implies that $h$ is independent of $x_i$.
In particular, we have that $X_{\kappa_i}$ and $Z_h$ act non-trivially on the Hilbert spaces of different parties, and we can write their product in a tensor product form.
\end{proof}

Due to this, it turns out that internally commuting stabilizers exist only for a subset of functions: 

\begin{prop}
\label{prop:internally_commuting}
Let $\left\{S_{f,\kappa_i}\right\}_{i=1}^n$ be a set of stabilizers which uniquely determines a state $\ket{f}$ and let the $\kappa_i=\pi_i^{-1}\kappa_i^{+}\pi_i$ be fixed $d$-cycles.
The stabilizers $S_{f,\kappa_i}=X_{\kappa_i}Z_{f\circ\kappa_{i} - f}$ commute internally if and only if the function~$f$~can~be~written~as 
\begin{equation}
  f=f^\prime \circ \pi_1\circ \dots \circ \pi_n,
\end{equation}
where $f^\prime$ has degree at most $1$ in each variable.
\end{prop}
The proof can be found in the Appendix~\ref{appendixA}. 
This result shows, that functions that have degree at most $1$ in each variable indeed have a special property that is lost in the case of general FFE states.  
Those are precisely LFP (and hence LU) equivalent to the qudit hypergraph states defined in Ref.~\cite{2017PhRvA..95e2340S} (see also table \ref{table:graphcomparisons} and the discussion above it). 
Thus, the latter have stabilizers with a simpler structure than those of general FFE states. 
Nevertheless, the above construction is still a direct generalization of the qubit stabilizers formalism and one can still recover a similar structure. 
For example, as an additional minor result we present in Appendix~\ref{app:FFEwithContST} a family of states having continuous unitary stabilizers, with a construction that generalizes that of Ref.~\cite{PhysRevA.79.042318,hypergraph2014}.

\section{Definition of Tensor-Edge Hypergraph (TEH)}
\label{sec:TEH}
\noindent
Now, let us return to polynomiality of the finite functions.
As mentioned before, Boolean functions can be encoded in \emph{graph} states.
Thus, we find a mapping from Boolean functions to \emph{simple} graphs.
However, this mapping turns out \emph{not} to hold for $d$-valued finite functions. 
In fact, the association to a graph is inherently connected to the polynomial representation of the function.
While in $\mathds{F}_d $ every function $f_d$ can be represented by a polynomial, this is \emph{not} the case in $\mathds Z_d$.
As an example, consider a function $f:\mathds{Z}_4\to \mathds{Z}_4$ and its image $\left(0,2,0,3\right)$. 
In order to find its polynomial representation, consider a polynomial over $\mathds{Z}_4$, which can be in general written as $p(x)= c_0+c_1 x + c_2 x^2 + c_3 x^3$.
Now, set $f(i)=p(i)$ to find the coefficients $c_i$:
\begin{align*}
    0 &= c_0&\pmod{4}
    \\
    2 & = c_0 + c_1 +c_2 +c_3&\pmod{4}
    \\
    0 & = c_0 + 2 c_1 + 4 c_2 + 8 c_3&\pmod{4}
    \\
    3 & =  c_0 + 3 c_1 + 9 c_2 + 27 c_3&\pmod{4}
\end{align*}
A quick calculation shows that; $c_0=0,c_1 = 0,c_2 = 2-c_3$ and finally $2c_3 =1$.
Clearly this last equation has no solution since $2$ has no multiplicative inverse in $\mathds Z_4$.
Thus, there exist functions in $\mathds Z_d$ which can \emph{not} be represented by a polynomial.
In this light, we will use term \emph{polynomial functions} for functions which have a polynomial representation. 
Generally, the polynomial representation of polynomial functions is \emph{not} unique.
In fact, many equivalent polynomials can be chosen  to represent the same polynomial function. 
The uniqueness of the polynomial representative can be recovered by choosing a suitable normal form of a polynomial, which typically restricts both the degree of the \emph{monomials} and the value of the coefficients.
Importantly, there exist \bl{polynomial normal forms} for any $\mathds{Z}_d$ and any number of variables $n$ \cite{Singmaster1974,Selezneva2017}.
Note, that graphs in general are not associated to the polynomial function directly, but to its polynomial representative.
Thus, without picking a normal form, many equivalent graphs can be associated to a single state.

In finite fields, every function is a polynomial function and there exists a canonical normal form.
Thus, a distinction of functions, polynomial functions and their polynomial representatives is unnecessary.
However, in our case, the choice of normal form has a direct influence on the sparsity of the tensor we associate to the edges.
We opt for a normal form which has many ``local''  monomials, i.e., terms whose corresponding state can be generated by local operations, and as few monomials present as possible. 
Our choice of normal form recovers the well known representation of polynomial functions for prime dimensions, i.e., the representation by polynomials with every variable of degree smaller than $d$, with coefficients as well smaller than $d$.
However, for non-prime dimensions these restriction do not suffice.
For a more detailed review of the normal form of polynomial functions over $\mathds{Z}_d$ we employ see \cref{AP:poly}.

\begin{remark}
Note, that it is possible to define a finite field as well for prime powers $d=p^k$.
The construction for prime powers is slightly more complicated: The finite field is \emph{not} equivalent to $\mathds{Z}_{p^k}$, like in the prime case, but there exists a finite field with $p^k$ elements unique up to isomorphism.
We will not concern ourselves with these constructions in this work, since we are more interested to understand the structure of finite encoding functions, considering the difference between finite fields and rings of integers (mod $d$). 
\end{remark}

Before we continue our investigation of FFE states, we like to define a subset of these states which can be associated with a new kind of graph we call tensor-edge hypergraph (TEH).
We observe that the association of a graph is directly connected to the polynomial representation of the finite function.
Thus, we want to encode polynomials into graphs; where the graph is a pair $G=(V,E)$, with $V=\{1,2,\ldots,n\}$ a set of vertices and $E$ is a set of edges and a polynomial $p$ is a function of the form $p\left(x_1,\dots,x_n\right) = \sum_{i_1,\dots,i_n} c_{i_1\dots i_n} x_1^{i_1}\dots x_n^{i_n}$, where each $x_1^{i_1}\dots x_n^{i_n}$ is called a monomial.
The main idea is to encode the variables in the vertices and the coefficients of each monomial in the edges.
Starting with the simplest example, a bi-variate monomial, we explain the restrictions of this encoding and define increasingly more complex edges to encode more complex polynomials in them.

\begin{figure}[h!]
    \centering
   \includegraphics[width = \textwidth]{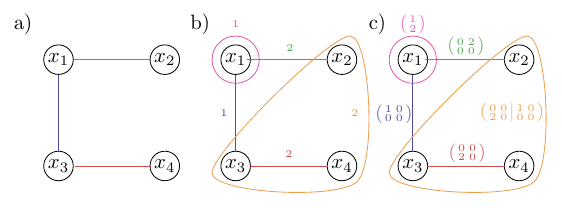}
  \caption{{Here are examples of graphical representations of polynomial functions. 
  Vertices correspond to variables of the depicted functions, (hyper-)edges differ based on the generality of description needed for a given function.}
   a) {A simple graph corresponding to a function} $p(x_1,x_2,x_3,x_4) =$ {\color{black!50!green!70}$x_1x_2$}+{\color{black!50!blue!70}$x_1x_3$}+{\color{black!30!red!70}$x_3x_4$}. {Each monomial of $p(x_1,x_2,x_3,x_4)$ corresponds to an edge between vertices.}
  b) {A hypergraph corresponding to a function} $p(x_1,x_2,x_3,x_4) =$ \textcolor{black!10!magenta!70}{$x_1$}+{\color{black!50!green!70}$2x_1x_2$}+{\color{black!50!blue!70}$x_1x_3$}+{\color{black!30!red!70}$2x_3x_4$}+{\color{black!10!orange!70}$2x_2x_3x_4$}. {Each monomial of $p(x_1,x_2,x_3,x_4)$ corresponds a hyperedge. 
  The coefficients of each {monomial} are represented as weight associated to each hyperedge.} 
  c) A TEH state corresponding to $p(x_1,x_2,x_3,x_4) =$ {\color{black!10!magenta!70}$x_1 + 2x_1^2$}+{\color{black!50!green!70}$2x_1x_2^2$}+{\color{black!50!blue!70}$x_1x_3$}+{\color{black!30!red!70}$2x_3^2x_4$}+{\color{black!10!orange!70}$2x_2^2x_3x_4 + x_2x_3x_4^2$}. {Here, multiple {monomials}  of $p(x_1,x_2,x_3,x_4)$ can correspond to the same hyperedge, since the hyperedges capture only information about which variables are present in a single {monomial}. 
  To also encode coefficients of the monomials present in $p(x_1,x_2,x_3,x_4)$ one needs to associate tensors $C^\alpha$ to each hyperedge $\alpha$. 
  Powers of each variable in monomial correspond to the list of indices in $C^\alpha$, while its coefficient value corresponds to the value of the corresponding tensor element. 
  Since the rank of the tensor depends on the number of nodes connected to the edge, in the case of the hyperedge connecting $x_2,x_3$ and $x_4$ (depicted in yellow), the corresponding tensor of rank $3$ is represented as two matrices. The left matrix corresponds to {monomials}  containing $x_4$, and the right matrix corresponds to monomials containing $x_4^2$}.}
    \label{fig:TEHEX}
\end{figure}

A given a polynomial $p$ is encoded into a graph by identifying each variable with a vertex and each edge with a non-zero coefficient of a monomial.
Consider the monomial $p(x_1,x_2)= x_1x_2$
This monomial is described by a \emph{simple} graph, with a edge between two vertices.
To see how this notion has to be expanded for  more complex polynomials, let us consider another polynomial $p(x_1,x_2,x_3)= x_1x_2+x_1x_2x_3$.
Clearly, a simple graph does not suffice because of  the second monomial in the polynomial.
However, by simply allowing the edges to be $n$-tuples incorporates monomials with $n$ variables. 
In this case, the graph is called a hypergraph and the edges are called hyperedges. 
Next, let us consider a polynomial of the form $p(x_1,x_2)= x_1x_2+2x_1x_2x_3$.
Again, we have to adjust the definition of the edges to be able to describe the different values of the coefficients.
To that end, a weight (or multiplicity) is added to each edge, that describes the coefficient of each monomial; these graphs are called weighed (hyper-)graphs.
Finally, consider the polynomial $p(x_1,x_2)= x_1x_2+x_1x_2^2$: 
To include monomials with higher exponents, again, the notion of the edge has to be extended.  
A tensor edge is a pair of a $n$-tuple of vertices $\alpha$, and a $n$-th order tensor $C^\alpha$ describing the coefficients of the monomials composed of variables in $\alpha$. 
Note, that the tensor will describe several monomials and not just a single one.
Thus, we need to collect the monomials which have the same variables, e.g., for a polynomial $p(x_1,x_2,x_3)= x_1x_2+x_1x_2^2+x_1x_2x_3+2x_1^2x_2x_3$ we would group the first two and the latter two terms together.
The monomials which are not present, i.e., their coefficient is zero,  correspond to zero entries in the tensor.
The physical motivation for this grouping is the fact that encoding monomials with the same set of variables with non-zero exponents requires interaction between the subsystems corresponding to these variables; see \cref{rm:modular_prep}. 
An illustrative example of these concepts is given in Figure \ref{fig:TEHEX}.
\leavevmode

With these \emph{tensor-edge} hypergraphs we are finally able to encode arbitrary polynomials with $c_{0\dots 0} =0$ into graphs.
In the light of the encoding of quantum states, this restriction is irrelevant since the  $c_{0\dots 0}$ corresponds to a global phase, i.e., $\omega^{p(\vec{x})}=\omega^{c_{0\dots0}+p'(\vec{x})}=\omega^{c_{0\dots0}}\omega^{p'(\vec{x})}$.
Although we allow ``local'' edges, i.e., edges that encode monomials with only a single variable, we like to point out that these can be avoided by LFP operations in the context of encoding of quantum states.
Let us  give a \emph{formal} definition of the tensor-edge hypergraphs. 
A $n$-vertex {\em tensor-edge hypergraph (TEH)} is a pair $G=(V,E)$, where $V=[n]$ is the set of vertices and $E$ is a set of tensor edges. 
A tensor edge is a pair $(\vec{\alpha},C^\vec{\alpha})$ where $\alpha\subset [n]$ and  $C^\alpha$ is a \bl{tensor}{} defined by grouping related monomials, 
i.e., we decompose a polynomial in the monomial basis as
\begin{align}
\label{genmonomial}                                                                     
p(\vec{x})
&= \sum_{\vec{e}} 
                c_{\vec{e}}
                \vec{x}^\vec{e} 
\end{align}                                                                                      
for some coefficients $c_{\vec{e}} \in \mathds{Z}_d$, $\vec{x}=(x_1,x_2,\ldots,x_n)$, $\vec{e}=(e_1,e_2,\ldots,e_n) \, \in \mathds Z_d^n$ and $\vec{x}^\vec{e} := x_1^{e_1},\dots,x_n^{e_n}$.

Now, we can  gather the terms in the monomial expansion~(\ref{genmonomial}) by terms that share the same variables with non-zero exponents. 
With this partitioning of terms, equation~(\ref{genmonomial}) becomes
\begin{equation}\label{tensoredgefndef}
  p(\vec{x})= \sum_{\alpha \subset[n]} \sum_{\small \substack{\vec{b}\in \mathbb{Z}_d^n:\\{\rm supp}(\vec{b})=\alpha}} C^\alpha_\vec{b}
       \vec x^{\vec b}
\end{equation}
where  $[n]=\{1,2,\ldots,n\}$ and we sum over all vectors with ${\rm supp}(\vec{b}) = \alpha$, i.e., all vectors which have~support~equal~to~$\alpha$. 

Clearly, TEHs are a generalization \bl{of simple}  graphs and hypergraphs.
Table~\ref{table:graphcomparisons} gives a summary of how various classes of \bl{simple} graphs, hypergraphs, and their generalizations, which we refer to collectively as graphs on $n$ vertices, are special cases of TEHs. 
Likewise, quantum states corresponding to the various classes of graphs are special cases of TEH states. 
 The table is organized by value of $d$ and the following restrictions:
  \begin{enumerate}
        \item [(1)] Even though $d>2$, all variables in all monomials in 
          $p$ are limited to exponents $0,1$. 
          \item [(2)] All monomials in the expansion of $p$  have at most two variables with
                nonzero exponents. 
  \end{enumerate}
  
  \begin{table}[h]
  \begin{center}
  \begin{tabular}{|c|c|c|c|}\hline
      graph/state type & restriction on $d$ & restriction on
    $p$ & references\\ \hline \hline
    TEH & none & none & (this paper)\\  \hline
    qudit hypergraph & none &  (1) & \cite{2017PhRvA..95e2340S} \\ \hline
      hypergraph & $d=2$ & none  & \cite{1367-2630-15-11-113022, 1404.6492, hypergraph2014}
      \\ \hline
      qudit graph & none & (1) and (2) & \cite{looi2008quantum,keet2010quantum} \\ \hline
      simple graph & $d=2$ & (2) & \cite{raussendorf2003measurement,hein2004multiparty}\\ \hline
  \end{tabular}
  \end{center}
  \caption{Comparison of various types of graphs and their generalizations and their corresponding quantum states.}
  \label{table:graphcomparisons}
  \end{table}

\begin{remark}
\label{rm:modular_prep}
Using this monomial expansion, we can recreate the ``simple'' recipe analogous to hypergraph states to prepare these TEH states, i.e., we can give a modular preparation scheme.
With this notation, the FFE state $\ket{p}$ that encodes the polynomial function $p=\sum_{\vec{e}}c_\vec{e}\vec{x}^{\vec{e}}$ can be also expanded in terms of monomial $Z$ operators:
\begin{align}
     \ket{p} = \frac{1}{\sqrt{d^n}}\sum_{\vec{x}\in (\mathds{Z}_d)^n} 
\omega_d^{p(\vec{x})} \ket{\vec{x}} = Z_p \ket{+}^{\otimes n} = \left(\prod_{\vec{e}}\left(Z_{\vec{x}^{\vec{e}}}\right)^{c_\vec{e}} \right)\ket{+}^{\otimes n} ,
\end{align}
where $\ket{+}=\frac{1}{\sqrt{d}}(\ket{0}+\ket{1}+\cdots+\ket{d-1})$  and $\vec{x}^{\vec{e}} = x_1^{e_1}x_2^{e_2}\cdots x_n^{e_n}$. 
We use the label $Z_{\vec{x}^{\vec{e}}}$ for the associated FP operation, which can be understood as a generalization of the qudit controlled-$Z$ gate.
Note, that the canonical qudit controlled-$Z$ corresponds to monomials with  $\vec{x}^{\vec{e}}$ with $\max(e_i)\leq 1$.
\end{remark}

\section{Local equivalence of Finite-Function-Encoding states}\label{sect:luequiv}
\noindent
In this section, we investigate the equivalence of finite-function-encoding states under LU and LFP operations.
Aside from the foundation's point of view, the question is important for understanding the potential of FFE states as resources for, e.g., quantum computation.  
Deciding whether multipartite quantum states are equivalent under local operations is one of the central questions of quantum information theory.
It has been studied for different subsets of quantum states and operations.
\textcolor{black}{The most prominent example is the paradigm of local operations and classical communications (LOCC) \cite{Nielsen:2011:QCQ:1972505,EltschkaSiewert2014,Chitambar_2014}}, \bl{which is motivated by an operational perspective.}{}
\bl{A simpler problem is the classification under local unitary (LU) operations, which however is still relevant for characterizing resources for quantum computation. }{}
Note, that the problem of characterizing \emph{all} LU equivalence classes  is notoriously difficult, even when the set of states considered is very restricted \cite{gross2007lulc,ji2008lulc}.
Already for the \textcolor{black}{comparably simpler} qubit hypergraph states a complete characterization of equivalence classes under local unitaries is an open problem\footnote{see
\url{https://oqp.iqoqi.univie.ac.at/local-equivalence-of-graph-states}}, and thus relaxed problems are investigated, e.g., \bl{equivalence under the restricted set of local so-called Clifford unitaries, i.e., unitaries that map Pauli matrices into Pauli matrices}{} \cite{Van_den_Nest_2005}. 
Another example is that of $k$-uniform states ~\cite{Goyeneche_2014,Goyeneche15,Goyeneche18,Huber18,Raissi20} \textcolor{black}{, i.e., $n-$partite states with all $k$-party \textcolor{black}{marginals} maximally mixed,} and their extremal case of so-called absolutely maximally entangled (AME) states \cite{Scott_2004,AME2,Goyeneche_2014,Goyeneche15,Goyeneche18,Raissi20}, \textcolor{black}{(i.e. $n$-party $k-$uniform states with $k =\lfloor \tfrac n 2 \rfloor$)}.
Given certain $n$ and  $d$, the classification of $k$-uniform states is a long-standing open problem.
Besides even deciding whether such states exist for given $n,d$ and $k$, it is also hard to decide whether, e.g., two AME states are LU inequivalent.

Here, in addition to equivalence under general LU operations, we study equivalence of FFE states under LFP operations, since these always map a FFE state into a FFE state. 
A similar classification problem arises in the theory of Hadamard matrices, where so-called Butson type Hadamard matrices, are classified up to operations that correspond to our LFP operations~\cite{TadejZyczkowski2006}. 
\textcolor{black}{Note, that even this very restricted classification is known to be very complex~\cite{Tadej2006,TadejZyczkowski2006,lampio2017orderly}}.
First, we show that the problem of identifying all LFP equivalence classes becomes quickly infeasible with increasing  dimension $d$ as well as the number of parties $n$ involved  by bounding the number of equivalence classes \bl{$\ell_{d,n}$} from below.
Second, we investigate in some detail the classification of bipartite states, where
we give a full LFP and LU classification for prime dimension $d=3$ and composite dimension $d=4$.
We observe, that LFP operations can transform TEH states into generic FFE states, i.e., they connect polynomial and non-polynomial function encoding non-trivially.
\textcolor{black}{Furthermore, we show that, in composite dimension, most of the equivalence classes do not contain a TEH state.}
Third, we identify all equivalence classes in $d=6$ which contain a TEH state. 
These specific studies showcase once more the difficulty to give a \emph{full} classification of local equivalences for FFE states.
\textcolor{black}{However, for many tasks a complete classification is not necessary and whether two \emph{particular} states are locally equivalent is more relevant.}
Deciding whether two states are LFP equivalent is a far easier task than the general classification. However, a simple brute-force algorithm still needs  $\mathcal{O}\left({d!}^n\right)$ steps.
In other words, even this simpler problem becomes infeasible already for a relatively small $n$ and $d$.
\bl{Still, this problem can be tackled with the help of LFP invariants, which we also discuss. 
This way, we can} investigate the structure of certain particular classes of
bipartite FFE states, which include maximally entangled states, and connect it to the theory of complex Hadamard matrices \bl{of Butson type}{}.

\subsection{The LFP classification problem}
\noindent
Here, we investigate the classification problem of FFE states under LFP operations.
We give a lower bound on the number of equivalence classes, showcasing that a general classification becomes quickly infeasible with the dimension and the number of parties.
Secondly, we introduce a set of LFP invariants which can aid in deciding if two particular states are non-equivalent under LFP operations.
\subsubsection{A lower bound on the number of LFP classes}
\noindent
Before we investigate the local Pauli equivalence of FFE states, let us introduce some useful quantities.
Recall that every finite function $f\colon \mathds Z_d^{n}\to \mathds Z_d$ is completely described by a tuple of its image, i.e. $f \leftrightarrow (f(\vec{e}))_{\vec{e}\in\mathds Z_d^{ n}}$. 
By giving this tuple some more structure we can define an \emph{image} tensor $M_f$ in the $\mathds Z_d$-module $\mathds Z_d^{\otimes n}$ (analogous of a vector space defined over the ring of integers $\mathds Z_d$).
\begin{align}
    M_f = \left(f(e_1,\dots,e_n)\right)_{ e_1,\dots,e_n \,\in \mathds Z_d}\label{eq:imageTensor}
\end{align}
This image tensor is intimately related to the coefficients of the FFE state: the state's coefficient tensor $T_f$ is obtained by taking the \emph{element wise} exponential function ${\rm EXP}$
\begin{align}
    T_f = {\rm EXP}{\left(\frac{2 \pi\mathit{i}}{d}M_f\right)} .
\end{align}
Clearly, both $M_f$ and $T_f $ describe the state completely.
To simplify the notation, we sometimes identify $M_f$ with $f$ when clear from the context.
Let us now discuss how $M_f$ and $T_f$ are transformed under the action of a LFP operation:
The action of a local $X_\pi$ gate is  a permutation of certain entries of $M_f$ and $T_f$.
However, since the permutation is local, only a subset of these entries can be changed. 
In fact, let us assume that a permutation acts on the first system, which exchanges the values of $0$ and $1$ and leaves other values unchanged.
We find that in $M_f$, elements $f(0,e_2,\dots,e_n)$ are exchanged with $f(1,e_2,\dots,e_n)$, while all $e_i$ stay the same.
From this, we can easily generalize and see that a local $X_\pi$ operation acts as a permutation of rows, or columns both in $M_f$ and $T_f$ if they are matrices, i.e., in case of $n=2$, and as a permutation of their higher-dimensional counterparts if they are higher-order tensors.
 A $Z_{h}$ operation is \emph{local} if the function $h(x_1,\dots,x_n)$ is univariate. 
 \bl{For example, consider $Z_h$ acting on a first qudit only, this  means} that it transforms $f(0,e_2,\dots,e_n)$ to $f(0,e_2,\dots,e_n)+h(0)$, $f(1,e_2,\dots,e_n)$ to $f(1,e_2,\dots,e_n)+h(1)$ and so forth.
In terms of $M_f$ this means that we add a constant term to certain rows or columns \bl{or their higher-dimensional counterparts.}{} 
For $T_f$ this translates into \bl{multiplication of}  the corresponding \bl{rows and columns by}{} constant phases. 
Note, that these operations are \bl{a subset of} operations which define equivalence classes of complex Hadamard matrices~\cite{TadejZyczkowski2006}.
\bl{In fact, we can define a normal form under LFP similarly as in Hadamard matrix theory, where the coefficient tensor of the state is such that}
\bl{\begin{equation}\label{eq:LGPnormalform}
    (T_f)_{i_1,0,\dots,0}=(T_f)_{0,i_2,\dots,0}   = \dots = (T_f)_{0,\dots,0,i_n}=1 .
\end{equation}}
\bl{Following the terminology of Hadamard matrices, we call this normal form {\it dephased}.
In the case of $n=2$ we call the submatrix $(T_f)_{i_1,i_2}$ for $i_1,i_2 \in \{1,\dots,d-1\}$ the {\it core}.}
Clearly, every FFE state can be transformed into its dephased form by a set of \emph{unique} local $Z_h$ operations.
Thus, requiring that the state is in its dephased form fixes the local $Z_h$ operation.
However, this leaves the ambiguity on the $X_\pi$ operations, i.e., there is not a single unique normal form. 
Furthermore, although making permutations of the non-zero entries of the normal form leads to LFP equivalent states,
one can also permute the $T_f$ out of the dephased form and then return to it with a different $Z_h$
operation, leading potentially to a new LFP equivalent state.
Thus, \bl{to find all states LFP equivalent to a given state}{}, e.g., in the case $n=2$ where $T_f$ is a matrix, we have to \bl{scan} all permutations of rows and columns, \bl{and then apply the corresponding $Z_h$ operations to bring  the state back into the dephased form}.
Using these simple preliminary observations, we can derive a conceptually-simple algorithm to find the orbits of states under LFP operations. 
Starting from a dephased state, we can apply permutations followed by a return to the dephased form. 
Once, all permutations have been applied, all equivalent states are found.
Unfortunately, the number of permutations increases rapidly with $d$ and $n$: a rough estimate of the complexity of the algorithm is $\mathcal O\left((d!)^n\right)$.
Because of this inherent complexity, the method \bl{becomes infeasible} already for very small values of $d$ and $n$. 
However, we are able to use it in order to divide all $3^9$ bipartite qutrit FFE states into $9$ LFP equivalence classes and all $4^{16}$ bipartite ququart FFE states into $807$ LFP classes, which are described in detail in Appendix~\ref{AP:qutritClasses}.

Furthermore, this method provides a lower bound on the number of LFP classes of $n$-partite FFE states in dimension $d$, since it gives an upper bound on the number of states in each LFP class. 
\begin{prop}
A lower bound on the number of LFP equivalence classes $\ell_{d,n}$ for a $n$-partite qudit FFE state is
\begin{align}
    \ell_{d,n}\geq \left\lceil\frac{d^{d^n - n(d-1) - 1}}{(d!)^n}\right\rceil
\end{align}
\end{prop}
\begin{proof}
As calculated above, each normalized LFP class can have at most $(d!)^n$ states, together with the fact that there are $d^{d^n - n(d-1) - 1}$ dephased $n$ partite $d$ dimensional states gives the lower bound on the number of classes.
\end{proof}
This is \bl{not} a tight bound, as the size of the classes is typically much smaller than $(d!)^n$. 
The case studies with $n=2$ and $d=3,4$ confirm this. Our lower bounds evaluate to $\ell_{3,2} \geq 3$ and $\ell_{4,2} \geq 456$, while the real number of classes are $\ell_{3,2} = 9$ and $\ell_{4,2} = 807$. 
The bound is useful to demonstrate that increasing the value of either $d$ or $n$ deems the full characterization impractical, as $\ell_{5,2}\geq 10596382$ and $\ell_{3,3} \geq 16142521$.

\subsubsection{Invariants under LFP unitaries}\label{sec:LFPinvariants}
\noindent
\bl{With a different approach, one can look for functions of $T_f$ or $M_f$, which are invariant under the action of LFP operations, to provide sufficient criteria for LFP inequivalence of particular states. }
Thus, given two FFE states, such invariants can be helpful in deciding whether they are \emph{not equivalent}. 

\bl{As a simple example of an LFP invariant, let us consider the sum of all elements of the image tensor $M_f$:
\begin{align}\label{eq:LGPinv1}
  S(f)=\sum_{x_1,\dots,x_n}f(x_1,\dots,x_n) \mod d .
\end{align}
The LFP invariance of $S(f)$ can be easily proven by considering that $X$-type operations only change the order of summation and $Z$-type operations acting on variable $x_i$ change $S(f)$ to $S(f+h)= S(f)+S(h)$, with  $h(\vec{x})$ a univariable function of an arbitrary variable $x_i$. 
Furthermore, $S(h) = 0 \mod d$, since for each $k\in \{0,\dots,d-1\}$ images  $h(x_1,\dots,x_{i-1},k,x_{i+1},\dots,x_n)$ have the same value for all
$x_1,\dots,x_{i-1},x_{i+1},\dots,x_n \in \{0,\dots,d-1\}$. 
This means that $S(h)$ can be written as $d^{n-1} \sum_{k\in \{0,\dots,d-1\}} \allowbreak h(0,\dots,0,k,0,\dots,0)$, which is  equal to $0$ modulo $d$.

A more elaborate LFP invariant can be defined using a set of certain sums in the following way. 
First one chooses an index $j$ and splits $S(f)$ into $d$ sums, $S_{x_j=k}(f)$, one for each possible value $k\in \{0,\dots,d-1\}$. 
Then, one considers the set of all such sums indexed by $j$, which is also invariant under LFPs. 
Formally, for each index $j\in \{ 1,\dots ,n\}$, we define the sets
\begin{equation}\label{eq:LGPinv23}
   I_j := \{S_{x_j=k}(f)\}_{k = 0}^{d-1} ,
\end{equation}
where 
\begin{align}
S_{x_j=k}(f) =\sum_{\forall i\neq j, x_i = 0}^{d-1} f(x_1,\dots,x_{j-1},k,x_{j+1},\dots,x_n). 
\end{align} 
Again, for each index $j$, the set $I_j$ above is invariant under LFP operations.
Similarly to $S(f)$ one can see that $X$-type operations on variable $x_i$, where $i\neq j$ only change the order of summands in each partial sum $S_{x_j=k}(f)$, while $X$-type operation on $x_j$ exchanges some $S_{x_j=k}(f)$ with $S_{x_j=k'}(f)$ within the set $I_j$. 
Concerning $Z$-type operations, one can use a similar argument as in the case of $S(f)$ and conclude that for each $k$ and $j$ the partial sum $S_{x_j=k}(f)$ after a $Z$-type LFP operation can be calculated as $S_{x_i=k}(f+h)$, where $h$ is univariate and thus $S_{x_i=k}(h) = 0 \mod d$. 
Note, that for bipartite systems (which is the application we consider below) $M_f$ is a matrix and correspondingly the invariant $S(f)$ is just the sum of all matrix elements, the invariant set $I_0$ contains sums of all elements in each matrix row, while $I_1$ contains sums of all elements in each matrix column.
Note that also in the theory of Hadamard matrices a similar invariant set has been defined (cf. Eq.~\eqref{eq:Lambdinv} in Appendix~\ref{app:hadamard}) and has been very helpful to distinguish equivalence classes~\cite{Haagerup} (see also Appendix~\ref{app:hadamard}). 
}

\subsection{Local unitary equivalence of bipartite FFE states}
\noindent
Now we focus our attention on the easiest non-trivial case, i.e, the bipartite scenario.
Two bipartite pure states are locally unitary equivalent, if and only if they have the same Schmidt decomposition, which can be obtained by performing a singular value decomposition of their coefficient matrices \cite{Nielsen:2011:QCQ:1972505}.
The singular value decomposition allows for a simple algorithm to decide whether two FFE states are LU equivalent, and  it can also provide a witness of LFP inequivalence. 
The converse is not true, and in fact we can explicitly observe that there are states that are LU equivalent but not LFP equivalent.
\bl{For example}, matrix transposition and complex conjugation on $T_f$  do not change the LU class, but sometimes map states into two different LFP classes. 

\begin{remark}\label{rmk:LUvsLGP}
A bipartite $d$-dimensional FFE states represented by coefficient matrices $T_f$, $T_f^T$ and $T_f^*$ are locally unitary equivalent, since they have all the same singular value decomposition.
\end{remark}

In the following, we investigate LFP equivalence for small dimensions and in particular the structure of the maximally entangled FFE states, identify  states with maximally entangled subspaces and finally give a full characterization of LFP and LU equivalence for dimension $d=3,4$ while characterizing all TEH states in $d=6$.
\subsubsection{Maximally entangled states}
\noindent
For general dimensions, we can start by looking at bipartite maximally entangled states. 
This is on the one hand interesting for potential practical applications of FFE states, and on the other hand it elucidates the difficulties of fully describing LFP classes of FFE states. 
Let us introduce some language: 
A $d \times d$ square matrix is called a {\it Butson type} Hadamard matrix  $H(d,q)$  if  $HH^\dagger=d \mathds{1}_d$ where all elements are $q$-th roots of unity and $\mathds{1}_d$ is the identity \bl{matrix} in dimension $d$. 
Then, let us recall that a bipartite pure state is maximally entangled whenever the marginals are maximally mixed. 
In our case, this means that a FFE state $\ket{f}$ is maximally entangled whenever  
 $\frac{1}{d}\sum_{k=0}^{d-1} \omega_d^{f(k,i)-f(k,j)} =  \delta_{i,j}$, where $\delta_{i,j}$ is the Kronecker delta. 
In other words, the (unnormalized) coefficient matrix of the state is unitary, i.e., $d T_fT_f^\dagger = \mathds{1}_d$ holds. 
Thus, maximally entangled FFE states correspond to Hadamard matrices, and more specifically to those of type $H(d,d)$.
An exemplary maximally entangled FFE state is precisely a traditional bipartite qudit graph state, namely the state that encodes the function $f_d(x,y)= x y$.
In this case, the coefficient matrix of the state is the finite Fourier transform $F_d$. 
Furthermore, all states obtained by applying LFP operations to the $f_d(x,y)= x y$ are also maximally entangled. 
In fact, Hadamard matrices themselves are also classified up to rows and columns permutations and multiplication by a diagonal matrix of complex phases.
From the point of view of finite functions, we see that in particular we can compose the above monomial as $\pi_1(x)\pi_2(y)$ with arbitrary permutations and still get a maximally entangled FFE state.

From the known results on complex Hadamard matrices, we are able to provide a characterization of the maximally entangled  FFE states  for low dimensions.
\bl{We also observe} that a full characterization for arbitrary $d$ cannot be given in simple terms, since the characterization of complex $H(d,d)$ matrices remains, despite decades of efforts, an open problem. 
So far, full characterization is given for $d=4$ and for (small) prime dimensions $p\leq 17$.
For those prime dimensions, it is known that the Fourier transform matrix represents the unique LFP class (corresponding \bl{to} maximally entangled FFE states \bl{in our language}~\cite{lampio2017orderly}).

The cases $d=4$ and $d=6$ are also instructive to report, since they are useful to clarify to some extent the additional structure that arises for \bl{composite} dimensions. 
For the case $d=4$, all Hadamard matrices of type $H(4,4)$ are well classified, and it is known that there exists a single continuous $1$-parameter family of them~\cite{TadejZyczkowski2006} (see also Appendix~\ref{app:hadamard}).
To fit in our definition of FFE states, we additionally require the entries to be only $4$-th root of unity, and making this additional restriction results in having exactly $2$ LFP inequivalent maximally entangled FFE states, which we label as $\ket{f_4}$ and $\ket{f_{2\times 2}}$, because 
the two corresponding coefficient matrices are respectively $\tfrac 1 4 F_4$ and $\tfrac 1 4 F_2\otimes F_2$.
Additionally, we can ask whether both those matrices have representative polynomial functions $p_f: \mathbb Z_4^2 \rightarrow \mathbb Z_4$, and it turns out that this is the case. 
The two functions are $f_4= x y$, and $f_{2\times 2}= x y^2+x^2y+2xy$. See also Table~\ref{table:d4classes}.

It is interesting to notice that these two TEH states are LU equivalent, but  not connected by any of the unitary transformations mentioned in Remark~\ref{rmk:LUvsLGP}. 
This follows from the fact, that there are two LFP inequivalent Hadamards of type $H(4,4)$, and the simple observation that $F_4 = (F_2 \otimes F_2 ) (F_2 \otimes F_2)^\dagger F_4$.
Thus, we can see that 
\begin{equation}\label{eq:specUF4F22}
   \ket{f_4}=\id_4 \otimes \left[F_4^T (F_2 \otimes F_2)^*\right] \ket{f_{2 \times 2}}  ,
\end{equation}
which means that two TEH states belonging to different LFP classes are connected by a particular local unitary, coming from a product of Hadamard matrices.
At the level of the coefficient matrices, it can be seen \bl{(cf.} Eq.~\eqref{eq:d4contfam} \bl{in Appendix \ref{app:hadamard})}{} that the operation connecting the two LFP classes is a Hadamard product with a particular matrix.
After making this simple observation, we can further notice that the same local unitary above maps other LFP inequivalent FFE states into each other. 
See also the results of our brute-force calculations in $d=4$ afterwards.

For $d=6$ the theory of Hadamard matrices becomes already extremely complicated and not all $H(6,6)$ matrices are known. 
For example, analogously to $d=4$, a continuous family is known  which includes $F_6$ and $F_3\otimes F_2$. 
These two matrices are however, contrary to the analogous $d=4$ case, LFP equivalent \cite{Tadej2006}. 
It is also curious to observe that the state with  \bl{the} coefficient matrix  $\tfrac 1 {6} F_3 \otimes F_2$ is not a TEH state.
In other words, the corresponding function $f_{3\times 2}$ (cf. Eq.~\eqref{eq:f32ands6}) is not a polynomial.
This can be observed by direct interpolation or also by listing all polynomials $p_f: \mathbb Z_6^2 \rightarrow \mathbb Z_6$~(see Table \ref{table:d6classes}).
It is also interesting to observe that there is a special matrix $S_6\in H(6,6)$ which does not belong to any continuous family and is also connected to other mathematical problems \footnote{It was in fact found by Tao in relation to the so-called Fuglede's conjecture. See Ref.~\cite{Tao2004} for further details.}. 
The function $s_6$ (cf. Eq.~\eqref{eq:f32ands6}) \bl{is} also not a polynomial.
In fact, we can see from Table~\ref{table:d6classes} that the only TEH maximally entangled state has a coefficient matrix given by $\tfrac 1 {6} F_6$.
As in the $d=4$ case, we can construct local unitaries that map FFE states in different LFP classes into each other. 
In this case $(F_3\otimes F_2)^\dagger F_6$ does not work since $\ket{f_6}$ and $\ket{f_{3\times 2}}$ are in the same LFP class~\cite{Tadej2006}. 
However, for example the \textcolor{black}{following local unitary mappings} 
\begin{equation}
\begin{aligned}
        \ket{s_6} =&\id_6 \otimes  \left[S_6^T (F_3\otimes F_2)^*\right] \ket{f_{3\times 2}}\\
        \ket{s_6} = &\id_6 \otimes S_6^T F_6^* \ket{f_{6}}
\end{aligned}
\end{equation}
do the job, \textcolor{black}{since $\ket{f_{3,2}}$ and $\ket{s_6}$ are in different LFP classes and so are $\ket{f_6}$ and $\ket{s_6}$}.
To conclude this discussion, we refer to \cite{TadejZyczkowski2006,lampio2017orderly} and references therein for further details about Butson type Hadamard matrices~\footnote{See also \url{https://chaos.if.uj.edu.pl/~karol/hadamard/} and \url{https://wiki.aalto.fi/display/Butson/Butson+Home} for an up-to-date catalog of complex Hadamard matrices}.

\subsubsection{States with low Schmidt rank}
Let us now focus on LFP classes with lower Schmidt rank.  
A natural question is whether it is possible to find bipartite states of dimension $d$ that have Schmidt rank $r$ and are maximally entangled in a subspace of dimension $r$.
In the case that $r$ divides $d$, it is easy to see that this is true by making a constructive proof that uses: (i) if $d=k \cdot r$ we can see the $d$-dimensional system is composed of a $k$ and a $r$-dimensional part; (ii) a $r$-th complex root of unity is also a $d$-th root of unity.
Then, one example of states with the required properties are those with coefficient matrices proportional to $F_r\otimes J_k$ and $J_k\otimes F_r$ respectively, where $J_k$ is the $k$-dimensional matrix with every coefficient equal to $1$. 
Building upon this idea, we can make a more general construction, that actually enables us to find several LU classes with non-maximal Schmidt rank. 
A first simple observation is the following:
\begin{lemma}\label{lemma:2}
FFE states with Schmidt rank smaller than $r$ can be obtained from representative functions of the type 
\begin{equation}\label{eq:rankbipfunc}
    f_r(x,y)=g_A(x) g_B(y) ,
\end{equation}
where either $g_A$ or $g_B$ is a function with only $r$ distinct outputs. 
\end{lemma}

\begin{proof}
Let us fix the function $g_A$ to have $r$ distinct outputs. 
Then, we have that the coefficient matrix $T_f$ has at most $r$ distinct rows. Thus, the rank of $T_f T^\dagger_f$ can be at most $r$, \bl{which is also the rank of the single-particle reduced density matrix of $\ket{f_r}$.} 
\end{proof}

\bl{It} is interesting to study more details on the LU and LFP classification in the case 
of $f_r(x,y)$ given by Eq.~\eqref{eq:rankbipfunc} with, say, $g_A$ having $r$ outputs and $g_B$ having $d$ outputs. 
\bl{In this case,  it is always possible to map the state to another one such that $g_B(y)=y$  using LFP unitaries.} 
Then, the coefficient matrix becomes proportional to a so-called {\it Vandermonde matrix}, which makes it easier to partially characterize the LU classification of the corresponding FFE states.
We make some observations in the following, corroborated by the discussion in Appendix~\ref{app:prooflemma2}. 

\begin{lemma}\label{lemma:ftimesf}
The state $\ket{f_r}$ corresponding to a function $f_r(x,y)$ as in Eq.~\eqref{eq:rankbipfunc} with $g_B$ having $d$ distinct outputs has exactly Schmidt rank  $r$. 
In the case of $r$ being a divisor of $d$ and each of the different $r$ outputs of $f_r$ appearing exactly $d/r$ times, $\ket{f_r}$ is maximally entangled in a $r$-dimensional subspace.
\end{lemma}

\begin{proof}
See Appendix~\ref{app:prooflemma2}.
\end{proof}

This construction uses precisely the ideas outlined above. In general, there are several $d\times d$ FFE states maximally entangled in a $r$-dimensional subspace whenever $r$ divides $d$. 
Obviously, all of them will be in the same LU class, but \bl{they can be in different LFP classes}.
In fact, some LFP classes can be distinguished by the invariant set \eqref{eq:LGPinv23}: \textcolor{black}{For example,} if we \textcolor{black}{consider odd dimensions $d$ and} take a function of the form $f(x,y)=g_A(x)y$ and we sum over the column index we get always zero, i.e., \bl{$I_1(f)=\{0,\dots,0\}$}, while if we take the ``transposed'' function $f^T(x,y)=x g_A(y)$ we get \bl{that $I_1(f^T)$ contains} nonzero elements whenever $\sum_y g_A(y)\neq 0$.
\textcolor{black}{Another} natural question is whether (and for which $d$) we can find some $\ket{f_r}$ that are also represented by a polynomial function. 
The answer is that at least one canonical rank-$r$ maximally entangled TEH state always exists, as we observe in the following:

\begin{prop}\label{prop:PFEinkdim}
Bipartite TEH states of dimension $d\times d$ that have Schmidt rank $r$ and are maximally entangled in a $r$-dimensional subspace exist whenever $r$ divides $d$.
One example is given by the state $\ket{m_{d/r}}$, where 
\begin{equation}
    m_{d/r}(x,y)=\frac d r xy .
\end{equation}
\end{prop}

\begin{proof}
The function $m_{d/r}(x,y)$
is a product of $g_A(x)=\tfrac d r x$ and $g_B(y)=y$, where $g_A$
has its outputs as the element of the subgroup $\mathbb Z_r \subset \mathbb Z_d$, i.e., it is an $r$-output function with each output appearing exactly $d/r$ times, due to the cyclicity of the function. 
The statement then follows directly from Lemma~\ref{lemma:ftimesf}.
\end{proof}

Note that the coefficient matrix of $\ket{m_{d/r}}$ given above is precisely
$T_{d/r}=\tfrac 1 d F_r\otimes J_{d/r}$,
which can be seen as the canonical form under LFP. 
Examples of such states for $d=6$ are given in Table~\ref{table:d6classes}. 

The particular case of $d=p^m$ being a prime power with $m>1$ is also special from the point of view of entanglement classes of FFE states. 
In that case, we can write any integer in $x\in \{0,\dots ,d-1\}$ in its $p$-ary expansion as $x=a_0 p^0+\dots +a_{m-1} p^{m-1}$, i.e.,
we can write any $d$-it as a $m$-tuple of $p$-its: $x\equiv (a_0,\dots,a_{m-1}):=\vec a$, where the $a_j$'s are in $\mathbb Z_p$.
Correspondingly, the single $d$-dimensional Hilbert space associated to $x$ can be viewed as a multi-qu$p$it space, i.e., as a tensor product $\mathcal H_p^{\otimes m}$, where $\mathcal H_p$ are $p$-dimensional. 
In turn, this fictitious multipartite structure leads to more richness also in the entanglement classes of FFE states. 
For example, we can observe in the following that more possibilities 
exist to construct TEH states that have Schmidt rank $r<d$ and are maximally entangled in the corresponding $r$-dimensional subspace.

\begin{prop}\label{prop:pdimTEH}
When $d=p^m$ with $m>1$, states associated to the functions 
\begin{equation}
\begin{aligned}
     m_{p^{m-1}}(x,y)&=x^{p^{m-1}} y , \\
     m^T_{p^{m-1}}(x,y)&= x y^{p^{m-1}} ,
    \end{aligned}
\end{equation}
are maximally entangled in a $p$-dimensional subspace. 
Furthermore, these two states belong to different LFP classes.
\end{prop}

\begin{proof}
The statement comes again from Lemma~\ref{lemma:ftimesf} and the fact that $m_{p^{m-1}}$ and $m_{p^{m-1}}^T$ are composed by a $p$-output function $g_A(x)=x^{p^{m-1}}$ and a $d$-output function $g_B(y)=y$. 
Furthermore, each of the $p$ outputs of $g_A(x)$ appears exactly $d/p=p^{m-1}$ times.  
To see this, let us consider the function $x^{p^{m-1}}$. 
We have that $(x+p)^{p^{m-1}}=\sum_{k=0}^{p^{m-1}}\binom{p^{m-1}}{k} x^{p^{m-1}-k}p^k$. 
It is a consequence of Kummer's Theorem (see~Lemma~\ref{lemma:combcoeffzeromodpm} in Appendix~\ref{app:prooflemma2}) that ${p^{m-1} \choose k}p^k\equiv 0 \mod p^m$ for $0<k$. 
Consequently, we have that $(x+p)^{p^{m-1}}=x^{p^{m-1}}$, meaning that the function $g_A(x)$ has a property of cyclicity of order $p$. 
To see that $g_A$ has $p$ distinct values, let $x,y$ be in the range $0\leq x,y\leq p-1$ and suppose that $g_A(x)=g_A(y)$. 
Then the difference $x^{p^{m-1}}-y^{p^{m-1}}$ is divisible by $p^m$.
Since $p$ is prime and $x,y<p$, it must be that $x=y$.
To distinguish the LFP classes, we can use the invariant \eqref{eq:LGPinv23}: 
We have \bl{$I_1(m_{p^{m-1}})=\{0,\dots,0\}$, while $I_1(m_{p^{m-1}}^T)$} contains nonzero elements as soon as $\sum_y y^{p^{m-1}}\neq 0$.
\end{proof}     
    
The construction in Lemma~\ref{lemma:ftimesf} does not work for constructing this type of low rank maximally entangled states
when the dimension $d=p$ is a prime number.
Still, the general construction in Lemma~\ref{lemma:2} contains actually far more low rank classes.
In particular, we have seen that single power functions $g_A(x)=c x^k$ in some cases have nice cyclic structures that have implications in the LU classification problem.
In Appendix~\ref{app:prooflemma2}, we provide a similar example for the case of prime dimensions, where a few rank-$2$ classes can be found in this way.
Thus, we have seen that the construction as in Lemmas~\ref{lemma:2} and \ref{lemma:ftimesf} provides FFE states for which the LU classification problem is essentially reduced to a combinatorial problem of studying the outputs of finite functions.
Furthermore, for functions associated to states belonging to the same LU class, one can use the LFP invariants of Sec.~\ref{sec:LFPinvariants} to make a (partial) LFP classification.

\subsubsection{Application of brute-force algorithms for small dimensions}
Here, let us summarize shortly the results of a general LU/LFP classification, which can be made for very small dimensions. 
First, for $d=2$ we can easily see that there are just two LU classes: the separable states, corresponding to functions of the form $f(x,y)=f_A(x)+f_B(y)$ and the maximally entangled states, corresponding to the function $f(x,y)=xy$.
For $d=3$ and $d=4$ it is also still feasible to use a brute-force algorithm to derive all LFP classes. 
Then, by performing the singular value decomposition of a representative matrix of each class, we can also find all the LU classes with this brute-force method. 
\bl{Another, slightly more efficient,  brute-force algorithm}  can be used to find all the LU classes in the bipartite case: 
It is sufficient to scan all possible $(d-1)\times (d-1)$ ``core'' matrices and calculate the traces of their powers up to the $d$-th. 
In this way, one can classify all possible characteristic polynomials of $T_f T_f^\dagger$, and thereby list all possible LU classes. 
This algorithm can provide a quicker answer to the problem in $d=3$ and $d=4$, but is still unfeasible for higher dimensions due to the extremely high number of core matrices to scan.
See appendix~\ref{app:brute-force} for more details and an explicit example in $d=3$. 

In the following, we summarize the results for small dimensions that also lead to statements valid in general.
The case $d=3$ is  exemplary of prime dimensions, and we know that in this case all finite functions are polynomials.
In Table~\ref{eq:d3classification} in Appendix~\ref{AP:qutritClasses} we summarize the list of LFP classes, grouped by the Schmidt rank.
Note the presence of classes as in Eq.~\eqref{eq:dpclassex}.
For the LU classes, we can easily observe that these $9$ LFP qutrit classes collapse into $6$ LU classes and that, in fact, the operations mapping LFP inequivalent states of the same LU class are just those mentioned in Remark~\ref{rmk:LUvsLGP}.

Next, the case of $d=4$ is particularly interesting because we can still fully solve it and see the additional complications that arise for non-prime dimensions, and at the same time the richness of structure that arises when the dimension is a prime-power. 
We find \bl{that} the number of LFP equivalence classes is $682$, while only $15$ of them have a \bl{TEH state in them}.
A summary of the LFP classes with a \bl{TEH state} representative, ordered by different Schmidt ranks, is presented in Table~\ref{table:d4classes} in Appendix~\ref{AP:ququartClasses}.
Furthermore, the number of LU equivalence classes is $127$ (cf. Table~\ref{table:FFEclasssum}) and only $7$ of them have a \bl{TEH state representative}. 
What is more interesting, is that now there exist LU operations different from those listed in Remark~\ref{rmk:LUvsLGP} that connect different LFP classes.
In fact, these LU operations are precisely those of the form $\mathds 1_4 \otimes \left[F_4^T (F_2\otimes F_2)^*\right]$ (plus eventually further LFPs), which however connect not just the maximally entangled states between each other, but also states in other LU classes. 
Another peculiarity of the $d=4$ case (also compared to the $d=6$) is that there is more than a single LFP class of TEH states which are maximally entangled in a lower dimensional subspace. 
Namely, besides the polynomial $m_2(x,y)=2xy$ which corresponds to a state maximally entangled in a $2$-dimensional subspace (as in Proposition~\ref{prop:PFEinkdim}), there are also $2$-dimensional maximally entangled states, with corresponding functions given by $x^2y$ and $xy^2$, precisely as in Proposition~\ref{prop:pdimTEH}.

For $d=5$ and $d=6$, it is already not possible to perform a full brute-force LFP/LU classification.
However, in the $d=6$ case it is possible to list all possible polynomial functions and thereby make a LFP/LU classification of TEH states, which is summarized in Table~\ref{table:d6classes} in Appendix~\ref{AP:qusixttClasses}.
Noticeable in this case \bl{is that}{} \bl{there exists} only a single maximally entangled class, corresponding to the function $xy$, and that the only lower-dimensional maximally entangled states are given by the construction as in Proposition~\ref{prop:PFEinkdim}, namely there is a $2$-dimensional maximally entangled state corresponding to the function $3xy$ and a $3$-dimensional maximally entangled state corresponding to the function $2xy$.
In Table~\ref{table:FFEclasssum} below, we summarize the characterization of the LU and LFP classes discussed above, for dimensions $d=3,4,5,6,7$.

Using these case studies, we can draw certain general conclusions concerning the structure of FFE states. 
The investigation of composite dimensions shows that polynomial and non-polynomial functions are related by LFP operations, and thus a simple characterization of the operations connecting all polynomial functions seems elusive. 
This problem is intimately connected to the question of polynomiality of finite functions over rings~\cite{Singmaster1974}. 
Furthermore, when considering LU operations, we can observe the number of unitaries, which collapse different LFP classes to a single LU class, increase with growing dimension.
While in the case of $d=3$ complex conjugation and transpose were sufficient to characterize all LU operations which connect LFP classes, in $d=4$ more LU operators were necessary. 
This implies that the structure of the LU operations which are necessary becomes increasingly difficult.
Finally, the connection with the theory of complex Hadamard \bl{matrices gives} access to a rich theory. 
However, many even basic properties of these matrices remain unknown and are the subject of ongoing research.
Seeing these complications already in the bipartite scenario hints at the increasing complexity of entanglement structures in the multipartite scenario.
On the other hand, any progress in the classification of Butson type Hadamard matrices will be directly translatable to results on FFE states.

  \begin{table}[h!]
  \begin{center}
  \begin{tabular}{|c|c|c|c|c|c|}\hline
      dim & $\#$ of LFP cl. & $\#$ of LU cl. & $\#$ of LFP ineq. MES & $\#$ LFP cl. for TEH & $\#$ LU cl. for TEH \\ \hline \hline
    2 & 2 & 2 & 1 & 2 & 2 \\  \hline
    3 & 9 & 6 & 1 & 9 & 6 \\ \hline
      4 &  807  & 127  & 2 & 17 & 7 
      \\ \hline
     5 & $>\left\lceil\frac{5^{14}}{576}\right\rceil$ & ? & 1 & $>\left\lceil\frac{5^{14}}{576}\right\rceil$ & ? \\ \hline
      6 & $>\left\lceil\frac{6^{21}}{400}\right\rceil$ & ? &  $\geq 4$ \cite{lampio2017orderly} & 27 & 12 \\ \hline
      7 & $>\left\lceil\frac{7^{34}}{518400}\right\rceil$ & ? & 1 & $>\left\lceil\frac{7^{34}}{518400}\right\rceil$ & ? \\ \hline
  \end{tabular}
  \end{center}
  \caption{In this table we summarize the number of LFP and LU  equivalence classes  for FFE and TEH states and give a number of LFP inequivalent maximally entangled states(MES). }
  \label{table:FFEclasssum}
  \end{table}

\section{Conclusions and outlook}
In this work, we introduced a framework that aims to exploit a potential interplay between high-dimensional logic
and quantum theory, both at a purely theoretical level and towards applications in quantum information. 
Motivated by generalizing the notion of qubit hypergraph state to the realm of high-dimensional quantum logic~\cite{Wang2020}, we define a set of quantum states which encode arbitrary multivariate, $d$-valued logical functions into their phases.
Naturally, the construction resembles very much that of qudit hypergraph states, previously introduced in Ref.~\cite{2017PhRvA..95e2340S}.
However, we took here the angle of arbitrary function-encodings, generalizing the construction of qubit hypergraph states, which encode all binary-logical functions, rather than that of generalizing the notion of controlled Pauli operations.
In fact, in our framework the natural generalization of Pauli operations is represented by elements of the generalized symmetric group, which we term finite-function-encoding Pauli operators, and are a much wider class of operations than the traditional qudit Heisenberg-Weyl group.
We observe in some detail how our notion generalizes that of \cite{2017PhRvA..95e2340S}  with several consequences. 
First, we point out that it is possible to associate certain graphical objects to states only when the encoded function is a polynomial. 
Such a generalized hypergraph, that we call tensor-edge hypergraph has a tensor attached to every edge, in order to have a one-to-one association with arbitrary polynomial terms.
Secondly, we also observe how the stabilizer group of FFE states is generally more complicated than that of qudit hypergraph states. 
We observe that the property of internal commutativity of stabilizers can be maintained only for the latter set of states (up to LFP equivalence). 
In the central part of our investigation, we studied the problem of LU classification of FFE states. 
Besides it being a traditionally relevant problem in quantum information, also related to the classification of resources
for applications, it showcases the complex structure of FFE states and their relation to the properties of functions over finite rings of integers.
Generalizing the idea of relating entanglement classes with (hyper)graph-theoretic properties (i.e., (hyper)edges and their possible multiplicities), we studied how entanglement classes of FFE states are associated to the properties of the underlying finite functions.  
This showed the interplay of our framework with important problems in combinatorics and number theory, like the classification of complex Hadamard matrices~\cite{TadejZyczkowski2006}.

Several open questions arise naturally from our study.
For example, finding the Clifford group associated to our finite-function-encoding Pauli operations seems pertinent
to extend our framework, as well as studying the associated local Clifford classification problem of FFE states.
In general, while our definitions and constructions work for any number of parties, the bulk of our present results are actually situated in the realm of bipartite states. 
While already here one can see interesting structures and identify open problems, we believe the most interesting road ahead concerns results on multipartite states. 
We expect that multipartite entanglement classification of FFE states will be, on the one hand, closely related to recent constructions of $k$-uniform states~\cite{Goyeneche18,Raissi20}, embedding those in a larger scenario containing far more structure. 
On the other hand, we expect that this will open up a plethora of consequent research towards applications of FFE states, for example for error correction, quantum algorithms and quantum computing in general.
In particular, we see measurement-based quantum computing as a promising potential application. 
In this scheme of quantum computation, the main resource is a multipartite entangled state, given in advance, and quantum gates are implemented via local measurements. 


Aside from these we expect that a deep investigation into multipartite FFE states will be strongly related with, and potentially have interesting implication for, several number theory problems, such as in particular, those related to complex Hadamard matrices and generalized permutation matrices.

Thus, 
it is very intriguing to explore further the connections between the mathematics of finite functions and the formalism of quantum theory with this angle, deepening further the applications of combinatorics and number theory to quantum information theory.
Similarly, perhaps in a bit of a more speculative perspective, one can think that progresses in implementations of FFE states in higher-dimensional quantum computation might find applications in solving complex combinatorial problems. 

In conclusion, the encoding of finite functions into higher dimensional quantum states yields a rich interplay of mathematical and physical tools. Our construction, despite its complexity, can still exploit a very rich and structured mathematical machinery to dig into the complex realm of multipartite, high-dimensional quantum states.
Furthermore, prominent abstract mathematical questions also gain physical relevance when applied to our framework.
On the one hand, our construction naturally embeds the theory of finite functions into quantum mechanics, 
thus relating combinatorial structures to properties of quantum states and their physical implementation. 
On the other hand, with concurrent developments in manipulating higher-dimensional quantum systems, it is of main practical importance to classify and distill relevant resources for different tasks.
In this sense, our construction also opens exciting explorations for potential practical applications, for example in the context of quantum computation, of a very abstract mathematical theory.

\section*{Acknowledgements}

We thank Mariami Gachechiladze, Stefan B{\"a}uml, Ludovico Lami and Otfried G\"uhne for discussions and Ferenc Sz\"oll\H{o}si for useful correspondence.
P.A., G.V. and M.H. acknowledge funding from the Austrian Science Fund (FWF)
through the START project Y879-N27, the Lise-Meitner project M 2462-N27 and the Zukunftskolleg project ZK 3. M.P. acknowledges the support of VEGA
project 2/0136/19 and GAMU project MUNI/G/1596/2019. D.L. acknowledges
funding from the National Science Foundation (NSF) through grants
PHY-1713868 and PHY-2011074. 

\bibliographystyle{unsrtnat}
\bibliography{main}
\appendix
\section{Polynomial Representability and Normal Form}\label{AP:poly}
We can make use of a normal form defined below to simply generate unique polynomials representing  all polynomial functions.
We will use the form given in \cite{Selezneva2017} since it favours ``local'' terms over ``non-local'' ones .
First let us introduce the \emph{composite degree}, a quantity which is used to restrict both \bl{the} degree of the monomials as well as the values of the coefficients of the unique polynomial representative we construct here.
The composite degree $c_{p,m}$ of a single variable monomial $x^e$ in a prime power ring $d=p^m$ is defined as the greatest number $t \in \{0,\dots,m-1\}$ such that the factorial $e!$ is divisible by $p^t$.
For multivariate monomials, $\vec{x}^\vec{e}$ with $x_i,e_i \in \mathds{Z}_d$ we have $c_{p,m}\left(\vec{x}^\vec{e}\right)= \min\left(m,\sum_{i=1}^n c_{p,m}(x_i^{e_i})\right)$. 
Now following Theorem 1 in \cite{Selezneva2017} we find:
\begin{remark}
For $d=p^m$ with $p$ a prime and $m>0$ any polynomial function $p\left(x_1,\dots,x_n\right)$ is in one-to-one correspondence with the polynomials of the form

\begin{align*}
    p\left(x_1,\dots, x_n\right)&= \sum_\vec{e}c_\vec{e}\vec{x}^\vec{e}
\end{align*}
with $c_{p,m}(\vec{x}^\vec{e})<m$, $c_\vec{e}\in \mathds{Z}_d^n$,  $c_\vec{e}<p^{m-c_{p,m}(\vec{x}^\vec{e})}$ 
\end{remark}

Let us  give an example for $d=p^m=2^2$ and $n=2$:
The first step is to identify the monomials which have a composite degree smaller than $m$:
Note that for a single variable monomials
\begin{equation*}
  c_{p,m}\left(x^e\right) 
 =
  \begin{cases}
  0 & \text{if $e \in \{0,\dots, p-1\} =\{0,1\}$}\\
  1& \text{if $e \in \{p,\dots, 2p-1\} = \{2,3\}$}
  \end{cases}
\end{equation*}
From this we can immediately see that all monomials of the form $x^iy^j$ with\bl{: (i) $i,j\leq 1$, (ii) $i\leq 1$ and $2\leq j \leq 3$, (iii) $j\leq 1$ and $2\leq i \leq 3$}  appear in the polynomial we are constructing. All others i.e. $2\leq i,j$ do not \bl{appear}.
Thus we find that any polynomial function over $\mathbb{Z}_4$ can be represented as:
\begin{align*}
    f(x,y) &= c_{00} +c_{10} x+c_{01}y+c_{11}xy 
    \\
    &\,+c_{20} x^2+c_{30} x^3+ c_{02}y^2+ c_{03}y^3+ c_{21}x^2y+c_{31}x^3 y+ c_{12}xy^2+c_{13}xy^3
\end{align*}
where the $c_{ij}$ are restricted by the composite degree of the corresponding monomials, i.e. the coefficients \bl{in} the first row are smaller than $2^{2-0}=4$ and the coefficients \bl{in} the second row are smaller than $2^{2-1}=2$.

Finally, let us take \bl{a look} at the case of \bl{arbitrary}{} composite dimension, i.e., $d=p_1^{m_1}\dots p_r^{m_r}$:
First, it is straightforward to use the method \bl{above} to find a unique polynomial representative for each prime factor $p_i^{m_i}$.
Now, each monomial of these polynomials will be found in the polynomial for the composite degree. If the same monomial appears in multiple polynomials for different $p_i^{m_i}$ its coefficients in the composite degree are found by the Chinese remainder theorem. 
In other words let $c_i\in \mathds{Z}_{p_i^{m_i}}$ be the coefficients of an arbitrary fixed monomial in each of the polynomials in prime factor $p_i^{m_i}$. Then the corresponding coefficient $c\in \mathds{Z}_d$ for the composite $d$ is the unique solution to the system of congruences  $c\equiv c_i\mod{(p_{i}^{m_i})} $. 

\section{Stabilizers: Proofs of the main Propositions}
\label{appendixA}

\begin{proof}[of \cref{cyclestabunique}]
Let $\ket{\psi}=\sum_{\vec{x}} c_{\vec{x}}\ket{\vec{x}}$ be any pure $n$-qudit state (we do not assume that $\ket{\psi}$ is a finite-function-encoding state, but we soon show that this must be the case), and assume that  $S_{f,\kappa_i}\ket{\psi}=\ket{\psi}$ for all $i$. Then we have
\begin{equation}\label{cyclicpermstabaction}
    X_{\kappa_i}Z_{f\circ \kappa_i-f}\ket{\psi} = \sum_{\vec{x}} \omega^{f(\kappa_i(\vec{x})) - f(\vec{x})}c_{\vec{x}} \ket{\kappa_i(\vec{x})}
= \sum_{\vec{x}} \omega^{f(\vec{x}) - f(\kappa_i^{-1}(\vec{x}))}c_{\kappa_i^{-1}(\vec{x})}  \ket{\vec{x}}.
\end{equation}
It follows that 
$$c_{\vec{x}}= \omega^{f(\vec{x}) - f(\kappa_i^{-1}(\vec{x}))}c_{\kappa_i^{-1}(\vec{x})}
$$
for all $i$ and for all $\vec{x}$. Because for each $i$ $\kappa_i$ is a $d$-cycle, letting $x_i$ vary, forces that coefficients
$$c_{x_1,x_2,\ldots,x_i,\ldots,x_n}, \qquad \text{and} \quad c_{x_1,x_2,\ldots,\kappa_i(x_i),\ldots,x_n}
$$
have the same norm, and also differ by a factor that is a $d$-th root of unity, for all $x_i$. 
Now allowing $\vec{x}$ to vary, we get that all the state vector coefficients $c_{\vec{x}}$ have the same norm and any two differ by a factor of a $d$-th root of unity. 
Thus we can always associate a finite function to the phases of the coefficients, and thereby establish that 
$\ket{\psi}$ is a FFE state.

Let us then denote the state as $\ket{g}$, calling $g$ the function encoded in its phase. 
With this notation Eq.~\eqref{cyclicpermstabaction} reads
\begin{equation}\label{cyclicpermstabaction2}
    X_{\kappa_i}Z_{f\circ \kappa_i-f}\ket{g} 
    = 
    \sum_{\vec{x}} 
        \omega^{f(\vec{x}) - f(\kappa_i^{-1}(\vec{x}))+g(\kappa_i^{-1}(\vec{x}))}
        \ket{\vec{x}}
\end{equation}
and we have
$$f(\vec{x}) - f(\kappa_i^{-1}(\vec{x}))+g(\kappa_i^{-1}(\vec{x}))=g(\vec{x})
$$
for all $i$ and for all $\vec{x}$. If we set $d(\vec{x})=f(\vec{x})-g(\vec{x})$, the above expression becomes
$$d(\vec{x})= d(\kappa_i^{-1}(\vec{x}))
$$
for all $i$ and for all $\vec{x}$. Once again, by varying $i$, then by varying $\vec{x}$, we conclude that $d$ is constant. 
This establishes that $\ket{g}$ is in fact equal to $\ket{f}$ (up to a global phase), and the proof of the proposition is complete.
\end{proof}

Before we prove 
\cref{prop:internally_commuting} let us state the following two remarks.

\begin{remark}
\label{lem:full_cycle_decomposition}
Two permutation cycles $\sigma= (i_1 i_2 \dots i_k)$ and 
$\kappa = (i^\prime_1 i^\prime_2 \dots i^\prime_k)$ of the same size $k$ are called \emph{conjugate} and one can write $\sigma = \pi^{-1}\kappa\pi$, where 
 $\pi(i^\prime_j)=i_j$ for $1\leq j\leq k$.
 In particular, any $d$-cycle $\kappa$ can be written as $\pi^{-1}\kappa^+\pi$, where $\kappa^+$ is a permutation mapping $x$ to $x+1\mod d$ and simultaneously $\pi^{-1}\kappa^+\pi$ is a $d$ cycle for any permutation $\pi$.
\end{remark}

\begin{remark}
Note also that, for a given $\kappa$, the choice of $\pi$ in Remark \ref{lem:full_cycle_decomposition} is not unique, since each $\kappa_i$ can be decomposed into $\pi_i^{-1}\kappa^{+}\pi_i$ in exactly $d$ different ways. This follows from the fact that vectors $(i_1 \dots i_d)$ and $(i_d i_1 \dots i_{d-1})$ represent the same $d$-cycle.
\end{remark}

Now we are finally ready to prove \cref{prop:internally_commuting}:
\begin{proof}[of \cref{prop:internally_commuting}]
\label{proof:internally_commuting}

For the first direction, we assume that the function $f$ can be decomposed as 
\begin{equation*}
    f=f^\prime \circ \pi_1\circ \dots \circ \pi_n ,
\end{equation*}
for some polynomial $f^\prime$ of degree at most $1$  in each variable. 
Then, let us consider the 
stabilizer $X_{\kappa_1}Z_{f\circ\kappa_1-f}$ where $\kappa_1$ is some $d$-cycle, which, by remark \ref{lem:full_cycle_decomposition} can be written as $\kappa_1 = \pi_1^{-1}\kappa^{+}\pi_1$.  Given that $f^\prime$ is of degree $1$ in $x_1$, 
we can write $f\circ\kappa_1$ as 
\begin{equation*}
  f\circ\kappa_1=  f^\prime(\pi_1(\kappa_1(x_1)),\pi_2(x_2),\dots,\pi_n(x_n)) = h_1(x_2,\dots,x_n) + \pi_1(\kappa_1(x_1)) g_1(x_2,\dots,x_n),
\end{equation*}
where for simplicity of notation we have incorporated all the 
other permutations $\pi_2,\dots , \pi_n$ into the functions $h_1$ and $g_1$.
Now using remark \ref{lem:full_cycle_decomposition} we consider the $d$-cycle $\kappa_1 = \pi_1^{-1}\kappa^{+}\pi_1$ and we have
\begin{equation*}
  f\circ\kappa_1 = h_1(x_2,\dots,x_n) + (\kappa^{+}\pi_1(x_1))g_1(x_2,\dots,x_n) ,
\end{equation*}
where $h_1(x_2,\dots,x_n)$ and $g_1(x_2,\dots,x_n)$ are functions linear in all variables.
In the next step we will use the fact that for every $d$ the permutation $\kappa^+(x)$ can be written as a polynomial $x+1 \mod d$, therefore the equation can be modified to
\begin{equation*}
  f\circ\kappa_1 = h_1(x_2,\dots,x_n) + (\pi_1(x_1)+1)g_1(x_2,\dots,x_n) ,
\end{equation*}
This, in turn, implies that 
\begin{equation*}
  f\circ\kappa_1 -f = g_1(x_2,\dots,x_n) ,
\end{equation*}
and from Lemma~\ref{lem:commutativity_cyclic_permutation} if follows that $X_{\kappa_1}Z_{f\circ\kappa_1-f}=Z_{f\circ\kappa_1-f}X_{\kappa_1}$. Repeating the same reasoning for all variables $x_i$ shows that we can find a set of internally commuting stabilizers $\{ X_{\kappa_i}Z_{f\circ\kappa_i-f}\}_{i=1}^n$ where all the $\kappa_i$ are $d$-cycles and thus \bl{by Proposition \ref{cyclestabunique}}{} they completely specify the state $\ket{f}$.

For the other direction, let us consider a set of stabilizers $\{ X_{\kappa_i}Z_{f\circ\kappa_i-f}\}_{i=1}^n$ where all the $\kappa_i$ are $d$-cycles. We first note that from Lemma~\ref{lem:commutativity_cyclic_permutation} it also follows that $X_{\kappa_i}Z_{f\circ\kappa_i-f}=Z_{f\circ\kappa_i-f}X_{\kappa_i}$ only if the function $f(x_1,\dots,\kappa_i(x_i),\dots,x_n)- f(x_1,\dots,x_n)$
does not depend on the variable $x_i$. In other words, taking $i=1$, we have
\begin{equation}\label{eq:pr2proof1}
f(\kappa_1(x_1),x_2,\dots , x_n) - f(x_1,\dots,x_n)= g_1(x_2,\dots , x_n) .
\end{equation}
From Remark~\ref{lem:full_cycle_decomposition} \bl{and the polynomial representation of $\kappa^+$}{} we also know that for every $d$-cycle $\kappa_1$ we can find some permutation $\pi_1$ such that
\begin{equation*}
    \kappa_1(x_1)=\pi_1^{-1}(\pi_1(x)+1) .
\end{equation*}
Thus, Eq.~(\ref{eq:pr2proof1}) simplifies to
\begin{equation*}
   f^\prime(y_1+1,x_2,\dots,x_n) - f^\prime(y_1,x_2,\dots, x_n)= g_1(x_2,\dots , x_n) ,
\end{equation*}
where we called $f^\prime=f\circ \pi_1^{-1}$ and $y_1=\pi(x_1)$. 
It is easy to see that the above difference equation in $y_1$ has solution
\begin{equation*}
    f^\prime(y_1,x_2,\dots,x_n)=h_1(x_2,\dots,x_n) + y_1 g_1(x_2,\dots , x_n) ,
\end{equation*}
where $h_1$ is a function independent of $x_1$. 
Hence, if Eq.~(\ref{eq:pr2proof1})
holds (for $i=1$), then the function $f$ can be written as
\begin{equation*}
f(x_1,\dots,x_n) = h_1(x_2,\dots,x_n) + \pi(x_1) g_1(x_2,\dots,x_n) . 
\end{equation*}
for some $g_1(x_2,\dots,x_n)$. 
Repeating the reasoning for all variables $x_i$ proves the Proposition.
\end{proof}

\section{FFE states with continuous unitary stabilizers}\label{app:FFEwithContST}

Besides the complete characterization of (discrete) stabilizer groups associated to general FFE states, following past works on hypergraph states, we can also ask for which FFE states continuous families of stabilizers exist.
In fact, classes of graph states and hypergraph states that have continuous local unitary stabilizers are known~\cite{PhysRevA.79.042318,hypergraph2014}.
In the following we generalize the results of Ref.~\cite{PhysRevA.79.042318,hypergraph2014} to the FFE scenario.

\begin{lemma}\label{lusybipsep}
Let $\ket{f}$ be a $n$-partite FFE state associated to a function $f({\bf x})$ such that
\begin{equation}
  f(\sigma(a),\sigma^{-1}(b),x_3,\dots,x_n) = f(\sigma(b),\sigma^{-1}(a),x_3,\dots,x_n)  ,
\end{equation}
for all $a,b \in \mathbb Z_d$ and for some permutation $\sigma$, 
and let $A$ be the operator such that $X_{\sigma}~=~\exp(i2\pi A/d)$.

The operator $S_{f,\sigma_1, \sigma_2^{-1}}(t)=\left[\exp(itA) \otimes \exp(-itA)\right]Z_{f\circ\sigma_1\circ \sigma_2^{-1}-f}$ stabilizes $\ket{f}$ for all values of $t$. 
 \end{lemma}
 
\begin{proof}First, we observe that $S_{f,\sigma_1,\sigma_2^{-1}}(t)$ is a stabilizer for $t=2\pi /d$. In fact, it is just the product of two operators as in Eq.~(\ref{def:stabilizer}), namely
\begin{equation}
    S_{f,\sigma_1,\sigma_2^{-1}}(2\pi/d) = X_{\sigma_1} X_{\sigma_2^{-1}} Z_{f\circ\sigma_1\circ \sigma_2^{-1}-f} = X_{\sigma_1}
    Z_{f\circ\sigma_1-f} X_{\sigma_2^{-1}}
    Z_{f\circ \sigma_2^{-1}-f} ,
\end{equation}
due to the commutation relation (\ref{XZ_commutation}) and the fact that $Z_{f\circ\sigma_1\circ \sigma_2^{-1}-f\circ \sigma_2^{-1}} Z_{f\circ \sigma_2^{-1}-f}=Z_{f\circ\sigma_1\circ \sigma_2^{-1}-f}$.
Then, for different values of $t \in \mathbb R$ we have that a sufficient condition for $S_{f,\sigma_1,\sigma_2}(t)\ket{f}=\ket{f}$ is that (cf.~Sec.~3 of~\cite{LyonsWalck2005})
\begin{equation}
    (A\otimes \mathds 1- \mathds 1 \otimes A)Z_{f\circ\sigma_1\circ \sigma_2^{-1}-f}\ket{f}=(A\otimes \mathds 1- \mathds 1 \otimes A)\ket{f\circ\sigma_1\circ \sigma_2^{-1}}=0 .
\end{equation}
This holds true from the assumption that $f(\sigma(a),\sigma^{-1}(b),x_3,\dots,x_n)=f(\sigma(b),\sigma^{-1}(a),x_3,\dots,x_n)$ for all $a,b \in \mathbb Z_d$, since in that case the state is invariant under exchange of the first two parties. 
\end{proof}
A particular example of function that fits in the above proposition and gives rise to a continuous family of stabilizers is $f({\bf x})=(x_1+x_2)g(x_3,\dots,x_n)$, with the corresponding permutation given by $\sigma=\kappa^+$ (or $\sigma=\kappa^-$). 
Note, how this function satisfies $f(\kappa^+(x_1),\kappa^-(x_2),x_3,\dots,x_n)=f(\kappa^-(x_1),\kappa^+(x_2),x_3,\dots,x_n)=f({\bf x})$, which is in turn invariant under exchanging $x_1 \leftrightarrow x_2$.

\section{Simple known results on Hadamard matrices $H(d,d)$ for $d=4$ and $d=6$}\label{app:hadamard}

In $d=4$, all Hadamard matrices of type $H(4,4)$ belong to a single continuous $1$-parameter family \cite{TadejZyczkowski2006}:
\begin{equation*}
  H_4(q) = \left(
    \begin{matrix}
    1 & 1 &1 &1 \\
    1 & i e^{iq} & -1 & -ie^{iq} \\
    1 & -1 & 1 & -1 \\
    1 & -ie^{iq} & -1 & ie^{iq} 
    \end{matrix}
    \right) ,
\end{equation*}
where $q$ is any real number.
This family can be also expressed as 
\begin{equation}\label{eq:d4contfam}
    H_4(q) = F_4 \circ_H {\rm EXP}(iq h_4(x,y)) ,
\end{equation}
with $h_4(x,y)=x^2y+xy^2+3xy \mod 4$, as in our language of finite functions, \bl{where EXP is the elementwise exponential}. 
Here the operation $\circ_H$ denotes the Hadamard product of two matrices, i.e., the elementwise multiplication.

For $d=6$ the theory of Hadamard matrices becomes already extremely complicated and not all $BH(6,6)$ matrices are known. 
For example, analogously to the $d=4$ a continous family with two real parameter $(a,b)$ is known:
\begin{equation*}
 H_6(a,b) = \left(
    \begin{matrix}
    1 & 1 & 1 & 1 & 1 & 1 \\
    1 & \omega_6 e^{ia} & \omega_6^2 e^{ib} & \omega_6^3 & \omega_6^4 e^{ia} & \omega_6^5 e^{ib} \\
    1 & \omega_6^2 & \omega_6^4 & 1 & \omega_6^2 & \omega_6^4 \\
    1 & \omega_6^3 e^{ia} & e^{ib} & \omega_6^3 & e^{ia} & \omega_6^3 e^{ib} \\
    1 & \omega_6^4 & \omega_6^2 & 1 & \omega_6^4 & \omega_6^2 \\
    1 & \omega_6^5 e^{ia} & \omega_6^4 e^{ib} & \omega_6^3 & \omega_6^2 e^{ia} &\omega_6 e^{ib} 
    \end{matrix}
    \right) .
\end{equation*}
\bl{This family} includes $F_6$ and $F_3\otimes F_2$. These two matrices are however, contrary to the analogous $d=4$ case, LFP equivalent \cite{Tadej2006}. 
Again, \bl{this family can be expressed as} a Hadamard product with $F_6$, namely
\begin{equation}
     H_6(a,b) = F_6 \circ_H {\rm EXP}(i (a g_6(x,y) + b h_6(x,y))) ,
\end{equation}
where $g_6(x,y)$ and $h_6(x,y)$ are, in our language, finite functions from $\mathbb Z_6^2$ to $\mathbb Z_6$.
Recently, a non-affine $4$-parameter family has been found~\cite{Szollosi_2012} and it has been conjectured that the full set of complex Hadamard matrices in $d=6$ consists of such a $4$-parameter family plus the isolated matrix $S_6$ found by Tao in~\cite{Tao2004}. 

It is also curious to observe that the functions corresponding to the matrices $F_3\otimes F_2$ and $S_6$, \bl{namely}
\begin{equation}\label{eq:f32ands6}
 f_{3,2} = \left(
    \begin{matrix}
    0 & 0 & 0 & 0 &0 &0 \\
    0 & 3 & 0 & 3 &0 &3 \\
    0 & 0 & 2 & 2 &2 &2 \\
    0 & 3 & 2 & 5 &4 &1 \\
    0 & 0 & 4 & 4 &2 &2 \\
    0 & 3 & 4 & 1 &2 &5 
    \end{matrix}
    \right) , \qquad \text{and} \quad 
    s_6 = \left(
    \begin{matrix}
    0 & 0 &0 &0 &0 &0 \\
    0 & 0 & 2 & 2 &4 &4 \\
    0 & 2 & 0 & 4 &4 &2 \\
    0 & 2 & 4 & 0 &2 &4 \\
    0 & 4 & 4 & 2 &0 &2 \\
    0 & 4 & 2 & 4 &2 &0 
    \end{matrix}
    \right) ,
\end{equation}
are not polynomials.
To conclude this section we remark that a set of invariants $\{ I_{a,b,c,d}(f) \}$ for equivalence classes of Hadamard matrices has been defined~\cite{Haagerup, TadejZyczkowski2006}, the elements of which in our context can be put in the form
\begin{equation}\label{eq:Lambdinv}
 I_{a,b,c,d}(f)  = f(a,b) - f(c,b) + f(c,d)- f(a,d) \quad {\rm for} \quad (a,b,c,d) \in \mathbb Z_d^{4} .
\end{equation}
Note that each of the elements of the set  is invariant under local $Z$-type operations since each index appears both with a plus and a minus sign. $X$-type operations, instead, can exchange elements between each other. However, the whole set remains invariant also under those transformations. 

\section{Details on LU and LFP classification of states as in Lemma~\ref{lemma:2}}\label{app:prooflemma2}

Here let us prove the statement in Lemma~\ref{lemma:ftimesf} and then make some further observations regarding the more general LU and LFP classification of states as in Lemma~\ref{lemma:2}.
Let us first introduce a notation for the single particle reduced density matrix of a bipartite FFE state:
\begin{equation}
    \rho_f := \tr_2(\ket{f}\bra{f}) =  \frac{1}{d^2}\sum_{a=0}^{d-1}\ket{f(\cdot,a)}\bra{f(\cdot,a)},
\end{equation}
where $\ket{f(\cdot,a)}$ is the (single particle) FFE state corresponding to $f(\cdot,a)=f(x,a)$ as a (single variable) function of $x$.
For states as in Lemma~\ref{lemma:2} we have:
\begin{equation}\label{eq:1partrholemma2}
    \rho_f= \frac{1}{d^2}\sum_{a=0}^{d-1}\sum_{x,y=0}^{d-1} \omega^{g_B(a)(g_A(x)-g_A(y))}\ket{x}\bra{y},
\end{equation}
which will be useful for the following observations.

\begin{proof}[of \cref{lemma:ftimesf}]
Let us consider 
a function $f_r(x,y)=g_A(x) g_B(y)$ such that $g_B$ has $d$ distinct outputs. 
Using LFP operations (i.e., permutations of the inputs) we can always transform this function into $g_B(y)=y$. 
Furthermore, we can also permute the inputs of $g_A$ such that the distinct $r$ outputs are precisely the first (i.e. $g_A(0)\neq g_A(1) \neq \dots \neq g_A(r-1)$).
Then, substituting $g_B(y)=y$ in Eq.~\eqref{eq:1partrholemma2} we can see that the single particle reduced density matrix
has elements which are either equal to $1/d$ or to zero \bl{due to the fact that $\sum_{i=0}^{d-1} \omega^i = 0$}. 
In particular, nonzero elements $(l,k)$ appear 
wherever $g_A(k)=g_A(l)$ and are symmetrical.
Thus overall, the matrix has elements $1/d$ along the whole diagonal and has some off-diagonal elements equal to $1/d$ appearing symmetrically. 
Hence, we can, e.g., subtract some rows from each other so to reduce the matrix to the form
\begin{equation*}
     \rho_f = \tfrac 1 d \left(
    \begin{matrix}
    \mathds{1}_r & \star \\
    \vec 0 & \vec 0 
    \end{matrix}
    \right),
\end{equation*}  
where we indicated as $\vec 0$ a matrix with all elements equal to $0$ and as $\star$ some leftover elements. This concludes the proof that indeed $\rho_f$ has rank $r$, which is by definition equal to the Schmidt rank of $\ket{f_r}$. 

For the second part of the proof, let us assume that the $r$ divides $d$ and that the outputs appear exactly $d/r$ times each. 
Then, we permute the rows of $T_f$ such that they repeat themselves exactly with cyclicity $d/r$ and we have 
\begin{equation*}
    \rho_f = \tfrac 1 d \id_r \otimes J_{d/r} ,\end{equation*}
where $\id_r$ is the identity in dimension $r$ and $J_{d/r}$ is the $(\tfrac d r)$-dimensional matrix with every coefficient equal to $1$.
Thus, we have that its eigenvalues are all equal to $1/r$.
\end{proof}

In what follows we state and prove a Lemma that supports
Proposition~\ref{prop:pdimTEH}.

\begin{lemma}\label{lemma:combcoeffzeromodpm}
  Let $p$ be a prime, let $m$ be a nonnegative integer, and
let $k$ be an integer in the range  $0< k \leq p^{m-1}$. We have
${p^{m-1}\choose k}p^k \equiv 0 \mod p^m$.
\end{lemma}

\begin{proof}
Kummer's Theorem says the following. Let $p$ be a prime, and let
$\nu(x)$ denote the highest power of $p$ that divides an integer
$x$. Given integers $a,b$  we have that 
$\nu({a\choose b})$ is the number of carries in the addition modulo $p$
of $b$ with $b-a$.

Let $a=p^{m-1}$ and let $b=k$. Then expansion of $b$ in powers of $p$ is
$$b=b_0p^0 + b_1p^1 + \cdots + b_{m-1}p^{m-1}.$$
Further, let $s$ be the smallest index such that $b_s\neq 0$.
If $s=m-1$, then $b=a$ and we have ${p^{m-1}\choose p^{m-1}}p^{p^{m-1}}
\equiv 0 \mod p^m$, so the Lemma holds.  If $s<m-1$, we have
$$a-b = (p-b_s)p^s + ((p-1)-b_{s+1})p^{s+1}+ ((p-1)-b_{s+2})p^{s+2} + \cdots +
((p-1)-b_{m-2})p^{m-2}.$$
When adding $b$ and $a-b$ modulo $p$, there is a carry when adding
in positions
$s,s+1,\ldots,m-2$, so that there are $m-s-1$ carries
total. Applying Kummer's Theorem, we have $\nu\big({a\choose b}\big) = m-s-1$. We
also have $b=k\geq p^s$, so $k>s$. It follows that
$$\nu\left({a\choose b}p^k\right)=m-s-1+k=m+(k-s-1)\geq m.$$ Thus the Lemma holds in this case,
and this concludes the proof.
\end{proof}

\subsection{Exemplary LU classification of states as in Lemma~\ref{lemma:ftimesf}}

Now, as an instructive example, let us try to make a classification of FFE states from the  construction in Lemma~\ref{lemma:ftimesf} for functions $g_A(x)$ with only two distinct outputs, which is essentially the simplest case.
Notice that when $g_A(x)$  is a constant function the FFE state belongs to the separable class.
In this case we have $\rho_f= J_{d}/d$ and there is only a single nonzero eigenvalue equal to $1$.
Thus, let us consider the case in which $g_A(x)$ has two distinct outputs.
In order to classify the corresponding FFE states into LU classes, we can calculate the characteristic polynomial $\chi(\rho_f)$ of $\rho_f$, which is of the form 
\begin{equation}\label{eq:ChpolT2out}
    \begin{aligned}
        \chi(\rho_f)(x)&= x^d+c_1 x^{d-1}+c_2 x^{d-2} , \quad \text{with} \\
        c_1 &=-\tr(\rho_f)=-1 , \\
        c_2 &=-\tfrac 1 2 \left(\tr(\rho_f^2)-\tr(\rho_f)^2\right) .
    \end{aligned}
\end{equation}
To calculate the coefficients above we have also to specify, for each output, how many inputs it corresponds to.
For example, let us assume that the first output only corresponds to one input and the other one to $d-1$ inputs.
In this case, we have that 
\begin{equation*}
 \begin{aligned}
  \rho_f &= \tfrac 1 d \left(
    \begin{matrix}
    1 & \vec 0 \\
    \vec 0^T & J_{d-1} 
    \end{matrix}
    \right) , 
    \end{aligned}
\end{equation*} 
with $\vec 0 = \left(0 , 0 , \dots , 0 \right)$.
Thus, in this case the full matrix $\rho_f$ has the eigenvalue $0$ with multiplicity $(d-2)$ and one eigenvalue $1/d$, which is the top left diagonal element.
Then, due to the fact that the total trace of the matrix is $1$, we deduce that the only remaining nonzero eigenvalue must be $1-1/d$.
Thus, the corresponding class of FFE states has~Schmidt~vector~given~by
\begin{equation}\label{eq:SchR2class1}
    \vec \lambda = (1/\sqrt{d},\sqrt{1-1/d}, 0,\dots,0) .
\end{equation}
As explained above, in more general cases, when the two distinct outputs correspond to different numbers of inputs, we essentially just need to calculate $\tr(\rho_f^2)$. 
From Eq.~\eqref{eq:1partrholemma2} we then obtain
\begin{equation}
\begin{aligned}
    \tr(\rho_f^2) &= \tfrac 1 {d^4} \sum_{ab} \sum_{xy} \omega^{(a-b)(g_A(x)-g_A(y))} = \tfrac{2d-1}{d^2} + \tfrac 1 {d^4} \sum_{a\neq b} \sum_{x\neq y} \omega^{(a-b)(g_A(x)-g_A(y))} \\
    &= \tfrac{2d-1}{d^2} + \tfrac{d-1}{d^3} (N_1^2-N_1+(d-N_1)^2-(d-N_1)) ,
\end{aligned}
\end{equation}
where we called $N_1$ the number of times output $o_1$ appears.

To conclude, let us briefly consider the even more general construction of Lemma~\ref{lemma:2}, still for the case in which $g_A(x)$ has just two distinct outputs and $g_B$ has an arbitrary number of outputs. 
In this case the Schmidt rank of the corresponding FFE state is still $2$, and thus, to make the LU classification we have still to calculate just $\tr(\rho_f^2)$. 
However, this time we have 
\begin{equation}
    \tr(\rho_f^2) = \tfrac 1 {d^4} \sum_{ab} \sum_{xy} \omega^{(g_A(x)-g_A(y))(g_B(a)-g_B(b))} ,
\end{equation}
and we see that this time the LU class depends also, besides the number of times they appear, on the concrete values of all the outputs of the functions $g_A$ and $g_B$.

\subsection{Exemplary classes constructed as in Lemma~\ref{lemma:2} for prime dimension}

In prime dimension $d$, we can use that $x^{d-1}=1$ for all $x>0$. Thus, the function $g_A(x)=k x^{d-1}=k$ has only two distinct outputs, namely $0$ and $k$, the latter appearing $d-1$ times.
Then, we can make the following combinations, which have associated rank-$2$ FFE states for every $1\leq k\leq d-1$:
\begin{equation}\label{eq:dpclassex}
\begin{aligned}
      f_{k,d-1} &= k x^{d-1}y , \\
      g_{k,d-1} &= k x y^{d-1} , \\
      h_{k,d-1} &= k x^{d-1}y^{d-1} .
\end{aligned}
\end{equation}
One can see that all the $\ket{f_{k,d-1}}$ and $\ket{g_{k,d-1}}$ belong to the same LU class, for which the two nonzero Schmidt eigenvalues are given by $\lambda_1= 1/\sqrt{d}$, and $\lambda_2=\sqrt{1-\lambda_1^2}$ (cf. Eq.~\eqref{eq:SchR2class1}). However, from the invariant \eqref{eq:LGPinv23} we see that $\ket{f_{k,d-1}}$ and $\ket{g_{k,d-1}}$ belong to different LFP classes. In fact, we have \bl{that $I_1(f_{k,d-1})$ has nonzero elements, while $I_{1}(g_{k,d-1})=\{0,\dots,0\}$}.

On the other hand, the $\ket{h_{k,d-1}}$ belong to different LU classes for $1\leq k\leq (d+1)/2$, while for the remaining $k$ the corresponding states are obtained via complex conjugation from the previous ones.
The LU classes are determined by calculating the characteristic polynomial of $\rho_h$, which is of the form given in Eq.~\eqref{eq:ChpolT2out} with $c_2=2\tfrac{(d-1)^2}{d^4}\cos(2\pi k/d)$.
Thus, we see that there is a different class for each 
$1\leq k\leq (d+1)/2$ and there is a symmetry under exchanging $k$ with $-k$.
On the other hand, for each $k$ they belong to a different LFP class, as it can be easily witnessed by the invariant (\ref{eq:LGPinv1}) which gives \bl{$S(h_{k,d-1})=k$}.

\section{Brute-force algorithms for LFP and bipartite LU classification}\label{app:brute-force}
As mentioned in the main text, we can derive a simple brute-force algorithm that finds all LFP classes of FFE states of a given number of parties $n$ and dimension $d$. 
Let us first describe it briefly and then show it in the practical example of the bipartite $3\times 3$ case. 
This algorithm makes use of what we called the ``dephased'' normal form under LFP, given by Eq.~(\ref{eq:LGPnormalform}).
Thus, it starts by considering a generic state in the dephased form, and its (initially empty) LFP class. 
Then, essentially, by scanning all elementary local finite-function-encoding Pauli operations, it fills all of its LFP class. Then it moves on to another (dephased) FFE state and repeats the procedure, until all possible \bl{dephased} matrices have been scanned and thus all possible LFP classes have been filled.

Explicitly, it goes as follows: 

\begin{algorithm}[H]
\SetAlgoNoLine
\SetAlgoNoEnd
\SetInd{0.5cm}{0.5cm}
Create an empty list of classes $Cl$\;
\For{all dephased coefficient tensor $T_f \in \mathbb C^{d^n}$}{
\If{$T_f \notin Cl$}
    {
    Create an empty list  $TMP$\;
     \For{all elementary permutation $X_\pi$}
     {
    Apply the permutation $T_f \longleftarrow X_\pi T_f X_\pi^\dagger$\;
    Transform it back to dephased form $T_f' \longleftarrow Z_{\rm deph f} T_f Z^\dagger_{\rm deph f}$ where $Z_{\rm deph f}=\bigotimes_{k=1}^n Z_{-f(x_k=0)}$\;
    Append $TMP$ with $T_f'$\;
    }
Append $CL$ with $TMP$
}
}
 \caption{Brute force LFP classification of $d^n$-dimensional FFE states}
\end{algorithm}

After we scan every possible dephased matrix for the given number of parties and dimension and repeat the above procedure we are left as the output with the full list of dephased FFE states for the number of parties and dimension considered, separated into the full list of LFP classes.

Let us now see how this algorith works in practice in the particular example of all $3\times 3$ FFE states. 
Let us start considering the function image
\[
f=\left(
\begin{matrix}
0&0&0\\ 
0&0&0\\
0&0&0\\
\end{matrix}
\right)  \longrightarrow T_f=\left(
\begin{matrix}
1&1&1\\ 
1&1&1\\
1&1&1\\
\end{matrix}
\right) 
\]
and apply the following list of passages:
\begin{enumerate}
    \item \label{item:1} Check if it belongs already to some LFP class. 
    \item \label{item:2} If not, store it to the next class and proceed further.
    \begin{enumerate}
    \item\label{step:swap} Swap two components of the coefficient tensor, e.g., 
\begin{equation}
    (T_f)_{1,i_2,\dots,i_n} \longleftrightarrow (T_f)_{2,i_2,\dots,i_n} ,
\end{equation}
    which amounts to apply a local $X_{\pi_1}$ to the FFE state. 
    In this case, our initial tensor is left invariant under this operation.
    \item\label{step:nform} Then, bring the tensor back to the dephased form, by subtracting the functions $f(0,i_2,\dots,i_n)$, $f(i_1,0,\dots,i_n)$ etc. (which amounts to apply
    local $Z$ operators of the form
        $Z_{-f(i_1=0)}$, $Z_{-f(i_2=0)}$, etc.).
        Again, in this case the tensor is already in dephased form. 
        \item Check if the tensor is already in some LFP class. 
        Yes, in this case it is already stored. 
        So, stop here and repeat from step \ref{step:swap} with a different swap. 

\begin{normalize}We see that our initial matrix is completely invariant under permutations of rows and columns. 
Thus it is the only member of its class, which is that of product states. 
In fact, this corresponds to the FFE state $\ket +^{\otimes 2}$.

Afterwards, we can consider for example the matrix (remember that we have to change only its core)
\[
f=\left(
\begin{matrix}
0&0&0\\ 
0&1&0\\
0&0&0\\
\end{matrix}
\right)  \longrightarrow T_f=\left(
\begin{matrix}
1&1&1\\ 
1&\omega_3&1\\
1&1&1\\
\end{matrix}
\right) ,
\]      
 and apply again the passages \ref{item:1} and \ref{item:2} above. 
 This time, permutations of the matrix rows and columns give a nontrivial result.
 Let us then swap columns $1$ and $3$. 
 We get (for simplicity working only with the function image matrix)
 \[
\left(
\begin{matrix}
0&0&0\\ 
0&1&0\\
0&0&0\\
\end{matrix}
\right) \stackrel{\mbox{cols} \ \  2 \leftrightarrow 3}{\longrightarrow} \left(
\begin{matrix}
0&0&0\\ 
0&0&1\\
0&0&0\\
\end{matrix}
\right) ,
\]
which is already in dephased form as well.
Then we do
\end{normalize}
 \item\label{step:storediffLGP} Store the resulting core state into the same LFP class. 

\end{enumerate} 
\end{enumerate}
\noindent
Then repeat from step \ref{step:swap} with a different swap.
We can now permute, for example, the rows $1$ and $2$ and columns $2$ and $3$, obtaining 
\[
\left(
\begin{matrix}
0&0&0\\ 
0&1&0\\
0&0&0\\
\end{matrix}
\right) 
\overset{\substack{
   \mbox{cols} \,  2 \leftrightarrow 3 \\
   \mbox{rows}\, 1 \leftrightarrow 2}
}{\longrightarrow}
\left(
\begin{matrix}
0&0&1\\ 
0&0&0\\
0&0&0\\
\end{matrix}
\right)
,
\]
which is now not in normal form. 
Thus, for bringing it back to normal form (step \ref{step:nform}) we subtract $1$ to all the elements of the third column, getting to (remember that we take numbers modulo $3$)
\[
\left(
\begin{matrix}
0&0&1\\ 
0&0&0\\
0&0&0\\
\end{matrix}
\right)
\stackrel{+2 \ \mbox{col} \ 3}{\longrightarrow}
\left(
\begin{matrix}
0&0&0\\ 
0&0&2\\
0&0&2\\
\end{matrix}
\right) ,
\]
which is a different dephased matrix belonging to the same LFP class. 
Thus, we store it to the same class and repeat with a different swap (step \ref{step:storediffLGP}). 

Let us for example swap the columns $1$ and $2$ and rows $1$ and $2$, obtaining 
\[
\left(
\begin{matrix}
0&0&0\\ 
0&1&0\\
0&0&0\\
\end{matrix}
\right) \overset{\substack{
   \mbox{cols} \,  1 \leftrightarrow 2 \\
   \mbox{rows}\, 1 \leftrightarrow 2}
}{\longrightarrow} \left(
\begin{matrix}
1&0&0\\ 
0&0&0\\
0&0&0\\
\end{matrix}
\right) ,
\]
which, again, can be brought back to dephased form by adding $2$ to the first column and then subtracting $2$ from the second and third rows. 
\[
\left(
\begin{matrix}
1&0&0\\ 
0&0&0\\
0&0&0\\
\end{matrix}
\right)
\stackrel{+2 \ \mbox{col} \ 1}{\longrightarrow}
\left(
\begin{matrix}
0&0&0\\ 
2&0&0\\
2&0&0\\
\end{matrix}
\right) ,
\]
This way we get to
\[
\left(
\begin{matrix}
0&0&0\\ 
2&0&0\\
2&0&0\\
\end{matrix}
\right) 
\stackrel{+1 \ \mbox{rows} \ 2,3}{\longrightarrow}
\left(
\begin{matrix}
0&0&0\\ 
0&1&1\\
0&1&1\\
\end{matrix}
\right)
,
\]
which is yet another dephased matrix belonging to the same LFP class. 
Proceeding further this way we will at some point find all possible matrices belonging to that class (which is listed as Class $2$ in Sec.~\ref{AP:qutritClasses}), exhausting also all possible permutations of the initially chosen state.
We can then pick another initial state which is not in the same class and repeat the same procedure, until we exhaust all possible dephased matrices.
This way it is possible to do by hand the full LFP classification in $d=3$, which results in the classes listed in Sec.~\ref{AP:qutritClasses} below. However, already the bipartite \bl{$d=5$} case and the tripartite $d=3$ case are too hard to be handled with regular computational power.

A brute force algorithm can be also given for finding all the LU classes in the bipartite case. The idea is based on the observation that each LU class in a bipartite $d\times d$ system is associated to a specific vector $\vec t_f=(\tr(\rho_f^2),\dots,\tr(\rho_f^d))$, where $\tr(\rho_f)=1$ has been omitted. I.e., all matrices $T_f$ in the same LU class will have the same values of the vector $\vec t_f$. This comes from the fact that all matrices $T_f$ in the same LU class will have the same vector of singular values $\vec \lambda(T_f)$, which, in turn, is related to the coefficients of the characteristic polynomial of $\rho_f$, and those coefficients are given by linear combinations of the elements of $\vec t_f$.  

Then, a brute force algorithm to find all the LU classes works by: (1) scanning all possible dephased coefficient matrices $T_f$; (2) calculating for each of them the value of the vector $t_f$ (3) store and count the different $t_f$ and the associated $T_f$. 

\begin{algorithm}[H]
\SetAlgoNoLine
\SetAlgoNoEnd
\SetInd{0.5cm}{0.5cm}
Create an empty list of classes $Cl$\;
Create an empty list of traces vectors $cl$\;
\For{all dephased coefficient matrices $T_f \in \mathbb C^{d^2}$}{
Calculate $\rho_f \longleftarrow T_f^\dagger T_f$\;
\For{$k$ from $2$ to $d$}{Create and calculate the components $t_f(k) \longleftarrow \tr(\rho_f^k)$\;}
\eIf{$t_f \notin cl$}{
Assign $t_f$ to $cl(end+1,end+1)\longleftarrow t_f$\;
Assign $T_f$ to the LU class $Cl(end+1,end+1)\longleftarrow T_f$ \;
}{Find the $i$ for which $t_f \in cl(i)$\;
Assign $T_f$ to the LU class $Cl(i,end+1)\longleftarrow T_f$ \;
}
}
 \caption{Brute force LU classification of $d^2$-dimensional FFE states}
\end{algorithm}

Note that also the performance of this algorithm scales very unfavorably with $d$, since all possible dephased $d\times d$ FFE states must be scanned. 
Still, in the $d=3$ it is easy to make the classification analytically and in the $d=4$ case in a short time with a regular pc. 

Let us then illustrate how the algorithm works in the $d=3$ case. 
Let us consider a generic $3\times 3$ dephased FFE state
\begin{equation}
    T_f = \tfrac 1 3 
\left(
\begin{matrix}
1 & 1 & 1 \\
1 & \omega_3^{a} & \omega_3^{b} \\
1 & \omega_3^{c} & \omega_3^{d} 
\end{matrix}
\right) .
\end{equation}
Let us now take the single particle reduced density matrix $\rho_f$
and calculate the traces of its $2$-nd and $3$-rd power. We obtain
\begin{equation}\label{eq:trrho23d3}
  \begin{aligned}
      \tr(\rho_f^2) &= \frac{5}{9} +
    \frac{4}{81} \bigg( R\omega_3 \left(a\right) +
    R\omega_3\left( b\right) +
    R\omega_3 \left(c\right)+
    R\omega_3\left( d \right)
    + R\omega_3\left( a-b\right) 
    + R\omega_3\left( a-c\right) 
    \\
    & + R\omega_3\left( b-d\right) + R\omega_3\left( c-d\right) + R\omega_3\left( a-b-c+d\right)
    \bigg) , \\
 \tr(\rho_f^3) &= \frac{29}{81} + \frac{16}{243} \bigg( 
 R\omega_3 \left(a\right) + R\omega_3\left( b\right) + R\omega_3 \left(c\right)+ R\omega_3\left( d \right) + + R\omega_3\left( a-b\right) + R\omega_3\left( a-c\right) 
  \\
    & + R\omega_3\left( b-d\right) + R\omega_3\left( c-d\right) + R\omega_3\left( a-b-c+d\right)
    \bigg)  \\
 &+\frac{2}{243} \bigg( R\omega_3\left( a-d\right)+R\omega_3\left(  b-c\right)+R\omega_3\left( a-b-c \right)+ R\omega_3\left(  a-b+d\right)+R\omega_3\left( a-c+d\right)
 \\
 &\phantom{+\frac{2}{243} \bigg(}+ R\omega_3\left(  b+c-d\right)  \bigg) ,
  \end{aligned}
\end{equation}
where we used the shortened notation $R\omega_3(x):=\cos(2\pi x/3)$.

Now, let us count how many different possible vectors $t_f$ there are in $d=3$. 
We see from Eq.~\eqref{eq:trrho23d3} that, since the function $R\omega_3$ is even in $x$, those expressions have several symmetries with respect to exchanging the numbers $(a,b,c,d)$.
For example, the exchanges $a\leftrightarrow d$ leaves both expression invariant, as well as $b\leftrightarrow c$.
The latter corresponds to just taking the transpose of $T_f$, while the former is connected with the complex conjugation of $T_f$.
Thus, we can consider just the cases in which $a\geq d$ and $b\geq c$, which are just $36$ out of the total $81$.
Also, the permutations $(a\leftrightarrow b, c \leftrightarrow d)$ 
and $(a\leftrightarrow c, b \leftrightarrow d)$
leave both expressions invariant.
Note that these are just LFP operations on the $T_f$ matrix (i.e., swaps of two rows or columns).
Due to these symmetries, we see that it is sufficient to scan just $21$ cases, of which only $6$ give 
different values of $\vec t_f$. 
Thus, we see that in total we have $6$ different LU classes in $d=3$.

This example also suggests that in general more clever algorithms can be found by identifying from the beginning what symmetries there have to be imposed on the matrix coefficients based, e.g., on LFP operations, transpositions and complex conjugations among other LU operations.
However, while the $d=4$ case is still easy to handle, and results in $127$ different LU classes,
larger dimensions become already intractable with regular computational power, even if symmetries are taken into account.
\newpage
\section{LFP and LU classes of bipartite qutrits}\label{AP:qutritClasses}
\noindent
In this appendix we list all the LFP equivalence classes of a bipartite qutrit FFE states. 
We added a \href{https://mfr.de-1.osf.io/render?url=https://osf.io/ycmrp/?direct\%26mode=render\%26action=download\%26mode=render} {comprehensive list of matrices} to an online repository.
\newcommand\Tstrut{\rule{0pt}{2.3ex}}

\begin{longtable}{ |c|c|l| }

\caption{The full classification of the equivalences of bipartite FFE/TEH states under LU and LFP operations in dimension $3$. 
The horizontal lines indicate different LFP classes while the singular values identify different LU classes. }\label{eq:d3classification}
\\
\hline
Schmidt Rank & Singular Values & Polynomial \\
\hline
\endfirsthead
\hline
Schmidt Rank & Singular Values & Polynomial \\
\hline
\endhead
\nopagebreak
\multirow{1}{4em}{$1$}&\multirow{1}{*}{{$\left(1.0,\,0.0,\,0.0\right)$}}&{$f(x,y)=  0$}\Tstrut
\\*
\hline
\nopagebreak
\multirow{9}{4em}{$2$}&\multirow{9}{*}{{$\left(0.90506,\,0.42527,\,0.0\right)$}}&{$f(x,y)=  x^{2} y^{2} + x^{2} y + x y^{2} + x y$}\Tstrut
\\*
\nopagebreak
&&{$f(x,y)=  x^{2} y^{2} + 2 \, x^{2} y + x y^{2} + 2 \, x y$}
\\*
\nopagebreak
&&{$f(x,y)=  x^{2} y^{2} + x^{2} y + 2 \, x y^{2} + 2 \, x y$}
\\*
\nopagebreak
&&{$f(x,y)=  x^{2} y^{2} + 2 \, x^{2} y + 2 \, x y^{2} + x y$}
\\*
\nopagebreak
&&{$f(x,y)=  x^{2} y^{2} + x y^{2}$}
\\*
\nopagebreak
&&{$f(x,y)=  x^{2} y^{2} + x^{2} y$}
\\*
\nopagebreak
&&{$f(x,y)=  x^{2} y^{2} + 2 \, x^{2} y$}
\\*
\nopagebreak
&&{$f(x,y)=  x^{2} y^{2} + 2 \, x y^{2}$}
\\*
\nopagebreak
&&{$f(x,y)=  x^{2} y^{2}$}
\\*
\hline
\nopagebreak
\multirow{9}{4em}{$2$}&\multirow{9}{*}{{$\left(0.90506,\,0.42527,\,0.0\right)$}}&{$f(x,y)=  2 \, x^{2} y^{2} + 2 \, x^{2} y + 2 \, x y^{2} + 2 \, x y$}\Tstrut
\\*
\nopagebreak
&&{$f(x,y)=  2 \, x^{2} y^{2} + x^{2} y + 2 \, x y^{2} + x y$}
\\*
\nopagebreak
&&{$f(x,y)=  2 \, x^{2} y^{2} + 2 \, x^{2} y + x y^{2} + x y$}
\\*
\nopagebreak
&&{$f(x,y)=  2 \, x^{2} y^{2} + x^{2} y + x y^{2} + 2 \, x y$}
\\*
\nopagebreak
&&{$f(x,y)=  2 \, x^{2} y^{2} + 2 \, x y^{2}$}
\\*
\nopagebreak
&&{$f(x,y)=  2 \, x^{2} y^{2} + 2 \, x^{2} y$}
\\*
\nopagebreak
&&{$f(x,y)=  2 \, x^{2} y^{2} + x^{2} y$}
\\*
\nopagebreak
&&{$f(x,y)=  2 \, x^{2} y^{2} + x y^{2}$}
\\*
\nopagebreak
&&{$f(x,y)=  2 \, x^{2} y^{2}$}
\\*
\hline
\nopagebreak
\multirow{6}{4em}{$2$}&\multirow{6}{*}{{$\left(0.8165,\,0.57735,\,0.0\right)$}}&{$f(x,y)=  2 \, x y^{2}$}\Tstrut
\\*
\nopagebreak
&&{$f(x,y)=  x y^{2}$}
\\*
\nopagebreak
&&{$f(x,y)=  x y^{2} + x y$}
\\*
\nopagebreak
&&{$f(x,y)=  2 \, x y^{2} + 2 \, x y$}
\\*
\nopagebreak
&&{$f(x,y)=  x y^{2} + 2 \, x y$}
\\*
\nopagebreak
&&{$f(x,y)=  2 \, x y^{2} + x y$}
\\*
\hline
\nopagebreak
\multirow{6}{4em}{$2$}&\multirow{6}{*}{{$\left(0.8165,\,0.57735,\,0.0\right)$}}&{$f(x,y)=  2 \, x^{2} y$}\Tstrut
\\*
\nopagebreak
&&{$f(x,y)=  x^{2} y$}
\\*
\nopagebreak
&&{$f(x,y)=  2 \, x^{2} y + x y$}
\\*
\nopagebreak
&&{$f(x,y)=  x^{2} y + 2 \, x y$}
\\*
\nopagebreak
&&{$f(x,y)=  2 \, x^{2} y + 2 \, x y$}
\\*
\nopagebreak
&&{$f(x,y)=  x^{2} y + x y$}
\\*
\hline
\nopagebreak
\multirow{18}{4em}{$3$}&\multirow{18}{*}{{$\left(0.77814,\,0.57735,\,0.24732\right)$}}&{$f(x,y)=  2 \, x^{2} y^{2} + 2 \, x y$}\Tstrut
\\*
\nopagebreak
&&{$f(x,y)=  2 \, x^{2} y^{2} + x y$}
\\*
\nopagebreak
&&{$f(x,y)=  2 \, x^{2} y^{2} + 2 \, x^{2} y + 2 \, x y$}
\\*
\nopagebreak
&&{$f(x,y)=  2 \, x^{2} y^{2} + x^{2} y + x y$}
\\*
\nopagebreak
&&{$f(x,y)=  2 \, x^{2} y^{2} + x^{2} y + 2 \, x y$}
\\*
\nopagebreak
&&{$f(x,y)=  2 \, x^{2} y^{2} + 2 \, x^{2} y + x y$}
\\*
\nopagebreak
&&{$f(x,y)=  2 \, x^{2} y^{2} + 2 \, x y^{2} + 2 \, x y$}
\\*
\nopagebreak
&&{$f(x,y)=  2 \, x^{2} y^{2} + x y^{2} + x y$}
\\*
\nopagebreak
&&{$f(x,y)=  2 \, x^{2} y^{2} + 2 \, x y^{2} + x y$}
\\*
\nopagebreak
&&{$f(x,y)=  2 \, x^{2} y^{2} + x y^{2} + 2 \, x y$}
\\*
\nopagebreak
&&{$f(x,y)=  2 \, x^{2} y^{2} + x^{2} y + x y^{2} + x y$}
\\*
\nopagebreak
&&{$f(x,y)=  2 \, x^{2} y^{2} + 2 \, x^{2} y + x y^{2} + 2 \, x y$}
\\*
\nopagebreak
&&{$f(x,y)=  2 \, x^{2} y^{2} + x^{2} y + 2 \, x y^{2} + 2 \, x y$}
\\*
\nopagebreak
&&{$f(x,y)=  2 \, x^{2} y^{2} + 2 \, x^{2} y + 2 \, x y^{2} + x y$}
\\*
\nopagebreak
&&{$f(x,y)=  2 \, x^{2} y^{2} + 2 \, x^{2} y + x y^{2}$}
\\*
\nopagebreak
&&{$f(x,y)=  2 \, x^{2} y^{2} + x^{2} y + 2 \, x y^{2}$}
\\*
\nopagebreak
&&{$f(x,y)=  2 \, x^{2} y^{2} + 2 \, x^{2} y + 2 \, x y^{2}$}
\\*
\nopagebreak
&&{$f(x,y)=  2 \, x^{2} y^{2} + x^{2} y + x y^{2}$}
\\*
\hline
\pagebreak[4]
\multirow{18}{4em}{$3$}&\multirow{18}{*}{{$\left(0.77814,\,0.57735,\,0.24732\right)$}}&{$f(x,y)=  x^{2} y^{2} + x y$}\Tstrut
\\*
\pagebreak[2]
&&{$f(x,y)=  x^{2} y^{2} + 2 \, x y$}
\\*
\nopagebreak
&&{$f(x,y)=  x^{2} y^{2} + x^{2} y + x y$}
\\*
\nopagebreak
&&{$f(x,y)=  x^{2} y^{2} + 2 \, x^{2} y + 2 \, x y$}
\\*
\nopagebreak
&&{$f(x,y)=  x^{2} y^{2} + 2 \, x^{2} y + x y$}
\\*
\nopagebreak
&&{$f(x,y)=  x^{2} y^{2} + x^{2} y + 2 \, x y$}
\\*
\nopagebreak
&&{$f(x,y)=  x^{2} y^{2} + x y^{2} + x y$}
\\*
\nopagebreak
&&{$f(x,y)=  x^{2} y^{2} + 2 \, x y^{2} + 2 \, x y$}
\\*
\nopagebreak
&&{$f(x,y)=  x^{2} y^{2} + x y^{2} + 2 \, x y$}
\\*
\nopagebreak
&&{$f(x,y)=  x^{2} y^{2} + 2 \, x y^{2} + x y$}
\\*
\nopagebreak
&&{$f(x,y)=  x^{2} y^{2} + 2 \, x^{2} y + 2 \, x y^{2} + 2 \, x y$}
\\*
\nopagebreak
&&{$f(x,y)=  x^{2} y^{2} + x^{2} y + 2 \, x y^{2} + x y$}
\\*
\nopagebreak
&&{$f(x,y)=  x^{2} y^{2} + 2 \, x^{2} y + x y^{2} + x y$}
\\*
\nopagebreak
&&{$f(x,y)=  x^{2} y^{2} + x^{2} y + x y^{2} + 2 \, x y$}
\\*
\nopagebreak
&&{$f(x,y)=  x^{2} y^{2} + x^{2} y + 2 \, x y^{2}$}
\\*
\nopagebreak
&&{$f(x,y)=  x^{2} y^{2} + 2 \, x^{2} y + x y^{2}$}
\\*
\nopagebreak
&&{$f(x,y)=  x^{2} y^{2} + x^{2} y + x y^{2}$}
\\*
\nopagebreak
&&{$f(x,y)=  x^{2} y^{2} + 2 \, x^{2} y + 2 \, x y^{2}$}
\\*
\hline
\nopagebreak
\multirow{12}{4em}{$3$}&\multirow{12}{*}{{$\left(0.84403,\,0.4491,\,0.29313\right)$}}&{$f(x,y)=  2 \, x^{2} y + 2 \, x y^{2}$}\Tstrut
\\*
\nopagebreak
&&{$f(x,y)=  x^{2} y + x y^{2}$}
\\*
\nopagebreak
&&{$f(x,y)=  x^{2} y + 2 \, x y^{2}$}
\\*
\nopagebreak
&&{$f(x,y)=  2 \, x^{2} y + x y^{2}$}
\\*
\nopagebreak
&&{$f(x,y)=  x^{2} y + x y^{2} + 2 \, x y$}
\\*
\nopagebreak
&&{$f(x,y)=  2 \, x^{2} y + x y^{2} + x y$}
\\*
\nopagebreak
&&{$f(x,y)=  x^{2} y + 2 \, x y^{2} + x y$}
\\*
\nopagebreak
&&{$f(x,y)=  2 \, x^{2} y + 2 \, x y^{2} + 2 \, x y$}
\\*
\nopagebreak
&&{$f(x,y)=  2 \, x^{2} y + 2 \, x y^{2} + x y$}
\\*
\nopagebreak
&&{$f(x,y)=  x^{2} y + 2 \, x y^{2} + 2 \, x y$}
\\*
\nopagebreak
&&{$f(x,y)=  2 \, x^{2} y + x y^{2} + 2 \, x y$}
\\*
\nopagebreak
&&{$f(x,y)=  x^{2} y + x y^{2} + x y$}
\\*
\hline
\nopagebreak
\multirow{2}{4em}{$3$}&\multirow{2}{*}{{$\left(0.57735,\,0.57735,\,0.57735\right)$}}&{$f(x,y)=  2 \, x y$}\Tstrut
\\*
\nopagebreak
&&{$f(x,y)=  x y$}
\\*
\hline
\end{longtable}

\section{LFP and LU classes of bipartite ququarts}\label{AP:ququartClasses}
\noindent
In this appendix we list all the LFP and LU equivalence classes of a bipartite qutquart \emph{TEH} states, i.e., only states which correspond to polynomial functions. 
We added a comprehensive list of the matrices  \href{https://mfr.de-1.osf.io/render?url=https://osf.io/jdpf7/?direct\%26mode=render\%26action=download\%26mode=render}{to an online repository}.
\begin{longtable}{ |c|c|l| }
\caption{The full classification of the equivalences of bipartite \emph{TEH} states under LU and LFP operations in dimension $4$. 
The horizontal lines indicate different LFP classes while the singular values identify different LU classes. \label{table:d4classes}}
\\
\hline
Schmidt Rank & Singular Values & Polynomial \\
\hline
\endfirsthead

\hline
Schmidt Rank & Singular Values & Polynomial \\
\hline
\endhead
\nopagebreak
\multirow{1}{4em}{$1$}&\multirow{1}{*}{{$\left(1.0,\,0.0,\,0.0,\,0.0\right)$}}&{$f(x,y)= 0$}\Tstrut
\\*
\hline
\nopagebreak
\multirow{2}{4em}{$2$}&\multirow{2}{*}{{$\left(0.92388,\,0.38268,\,0.0,\,0.0\right)$}}&{$f(x,y)= x^{2} y + x y^{2} + x y$}\Tstrut
\\*
\nopagebreak
&&{$f(x,y)= x^{2} y + x y^{2} + 3 \, x y$}
\\*
\hline
\nopagebreak
\multirow{4}{4em}{$2$}&\multirow{4}{*}{{$\left(0.86603,\,0.5,\,0.0,\,0.0\right)$}}&{$f(x,y)= x y^{3} + x y^{2}$}\Tstrut
\\*
\nopagebreak
&&{$f(x,y)= x y^{3} + x y$}
\\*
\nopagebreak
&&{$f(x,y)= x y^{3} + x y^{2} + 2 \, x y$}
\\*
\nopagebreak
&&{$f(x,y)= x y^{3} + 3 \, x y$}
\\*
\hline
\nopagebreak
\multirow{4}{4em}{$2$}&\multirow{4}{*}{{$\left(0.86603,\,0.5,\,0.0,\,0.0\right)$}}&{$f(x,y)= x^{3} y + x^{2} y$}\Tstrut
\\*
\nopagebreak
&&{$f(x,y)= x^{3} y + x y$}
\\*
\nopagebreak
&&{$f(x,y)= x^{3} y + x^{2} y + 2 \, x y$}
\\*
\nopagebreak
&&{$f(x,y)= x^{3} y + 3 \, x y$}
\\*
\hline
\pagebreak[4]
\multirow{4}{4em}{$2$}&\multirow{4}{*}{{$\left(0.86603,\,0.5,\,0.0,\,0.0\right)$}}&{$f(x,y)= x y^{3} + x^{2} y + x y^{2}$}\Tstrut
\\*
\nopagebreak
&&{$f(x,y)= x y^{3} + x^{2} y + x y$}
\\*
\nopagebreak
&&{$f(x,y)= x y^{3} + x^{2} y + x y^{2} + 2 \, x y$}
\\*
\nopagebreak
&&{$f(x,y)= x y^{3} + x^{2} y + 3 \, x y$}
\\*
\hline
\nopagebreak
\multirow{4}{4em}{$2$}&\multirow{4}{*}{{$\left(0.86603,\,0.5,\,0.0,\,0.0\right)$}}&{$f(x,y)= x^{3} y + x^{2} y + x y^{2}$}\Tstrut
\\*
\nopagebreak
&&{$f(x,y)= x^{3} y + x y^{2} + x y$}
\\*
\nopagebreak
&&{$f(x,y)= x^{3} y + x^{2} y + x y^{2} + 2 \, x y$}
\\*
\nopagebreak
&&{$f(x,y)= x^{3} y + x y^{2} + 3 \, x y$}
\\*
\hline
\nopagebreak
\multirow{5}{4em}{$2$}&\multirow{5}{*}{{$\left(0.70711,\,0.70711,\,0.0,\,0.0\right)$}}&{$f(x,y)= x^{2} y + x y$}\Tstrut
\\*
\nopagebreak
&&{$f(x,y)= x y^{2} + x y$}
\\*
\nopagebreak
&&{$f(x,y)= 2 \, x y$}
\\*
\nopagebreak
&&{$f(x,y)= x^{2} y + 3 \, x y$}
\\*
\nopagebreak
&&{$f(x,y)= x y^{2} + 3 \, x y$}
\\*
\hline
\nopagebreak
\multirow{2}{4em}{$2$}&\multirow{2}{*}{{$\left(0.70711,\,0.70711,\,0.0,\,0.0\right)$}}&{$f(x,y)= x y^{2}$}\Tstrut
\\*
\nopagebreak
&&{$f(x,y)= x y^{2} + 2 \, x y$}
\\*
\hline
\nopagebreak
\multirow{2}{4em}{$2$}&\multirow{2}{*}{{$\left(0.70711,\,0.70711,\,0.0,\,0.0\right)$}}&{$f(x,y)= x^{2} y$}\Tstrut
\\*
\nopagebreak
&&{$f(x,y)= x^{2} y + 2 \, x y$}
\\*
\hline
\nopagebreak
\multirow{8}{4em}{$3$}&\multirow{8}{*}{{$\left(0.80902,\,0.5,\,0.30902,\,0.0\right)$}}&{$f(x,y)= x^{3} y + x y^{3}$}\Tstrut
\\*
\nopagebreak
&&{$f(x,y)= x^{3} y + x y^{3} + x^{2} y + x y^{2}$}
\\*
\nopagebreak
&&{$f(x,y)= x^{3} y + x y^{3} + x^{2} y + x y$}
\\*
\nopagebreak
&&{$f(x,y)= x^{3} y + x y^{3} + x y^{2} + x y$}
\\*
\nopagebreak
&&{$f(x,y)= x^{3} y + x y^{3} + 2 \, x y$}
\\*
\nopagebreak
&&{$f(x,y)= x^{3} y + x y^{3} + x^{2} y + x y^{2} + 2 \, x y$}
\\*
\nopagebreak
&&{$f(x,y)= x^{3} y + x y^{3} + x^{2} y + 3 \, x y$}
\\*
\nopagebreak
&&{$f(x,y)= x^{3} y + x y^{3} + x y^{2} + 3 \, x y$}
\\*
\hline
\nopagebreak
\multirow{8}{4em}{$3$}&\multirow{8}{*}{{$\left(0.80902,\,0.5,\,0.30902,\,0.0\right)$}}&{$f(x,y)= x^{3} y + x y^{3} + x^{2} y$}\Tstrut
\\*
\nopagebreak
&&{$f(x,y)= x^{3} y + x y^{3} + x y^{2}$}
\\*
\nopagebreak
&&{$f(x,y)= x^{3} y + x y^{3} + x y$}
\\*
\nopagebreak
&&{$f(x,y)= x^{3} y + x y^{3} + x^{2} y + x y^{2} + x y$}
\\*
\nopagebreak
&&{$f(x,y)= x^{3} y + x y^{3} + x^{2} y + 2 \, x y$}
\\*
\nopagebreak
&&{$f(x,y)= x^{3} y + x y^{3} + x y^{2} + 2 \, x y$}
\\*
\nopagebreak
&&{$f(x,y)= x^{3} y + x y^{3} + 3 \, x y$}
\\*
\nopagebreak
&&{$f(x,y)= x^{3} y + x y^{3} + x^{2} y + x y^{2} + 3 \, x y$}
\\*
\hline
\nopagebreak
\multirow{4}{4em}{$3$}&\multirow{4}{*}{{$\left(0.70711,\,0.5,\,0.5,\,0.0\right)$}}&{$f(x,y)= x y^{3} + x^{2} y$}\Tstrut
\\*
\nopagebreak
&&{$f(x,y)= x y^{3} + x^{2} y + x y^{2} + x y$}
\\*
\nopagebreak
&&{$f(x,y)= x y^{3} + x^{2} y + 2 \, x y$}
\\*
\nopagebreak
&&{$f(x,y)= x y^{3} + x^{2} y + x y^{2} + 3 \, x y$}
\\*
\hline
\nopagebreak
\multirow{4}{4em}{$3$}&\multirow{4}{*}{{$\left(0.70711,\,0.5,\,0.5,\,0.0\right)$}}&{$f(x,y)= x y^{3}$}\Tstrut
\\*
\nopagebreak
&&{$f(x,y)= x y^{3} + x y^{2} + x y$}
\\*
\nopagebreak
&&{$f(x,y)= x y^{3} + 2 \, x y$}
\\*
\nopagebreak
&&{$f(x,y)= x y^{3} + x y^{2} + 3 \, x y$}
\\*
\hline
\nopagebreak
\multirow{4}{4em}{$3$}&\multirow{4}{*}{{$\left(0.70711,\,0.5,\,0.5,\,0.0\right)$}}&{$f(x,y)= x^{3} y + x y^{2}$}\Tstrut
\\*
\nopagebreak
&&{$f(x,y)= x^{3} y + x^{2} y + x y^{2} + x y$}
\\*
\nopagebreak
&&{$f(x,y)= x^{3} y + x y^{2} + 2 \, x y$}
\\*
\nopagebreak
&&{$f(x,y)= x^{3} y + x^{2} y + x y^{2} + 3 \, x y$}
\\*
\hline
\nopagebreak
\multirow{4}{4em}{$3$}&\multirow{4}{*}{{$\left(0.70711,\,0.5,\,0.5,\,0.0\right)$}}&{$f(x,y)= x^{3} y$}\Tstrut
\\*
\nopagebreak
&&{$f(x,y)= x^{3} y + x^{2} y + x y$}
\\*
\nopagebreak
&&{$f(x,y)= x^{3} y + 2 \, x y$}
\\*
\nopagebreak
&&{$f(x,y)= x^{3} y + x^{2} y + 3 \, x y$}
\\*
\hline
\nopagebreak
\multirow{2}{4em}{$4$}&\multirow{2}{*}{{$\left(0.5,\,0.5,\,0.5,\,0.5\right)$}}&{$f(x,y)= x^{2} y + x y^{2}$}\Tstrut
\\*
\nopagebreak
&&{$f(x,y)= x^{2} y + x y^{2} + 2 \, x y$}
\\*
\hline
\nopagebreak
\multirow{2}{4em}{$4$}&\multirow{2}{*}{{$\left(0.5,\,0.5,\,0.5,\,0.5\right)$}}&{$f(x,y)= x y$}\Tstrut
\\*
\nopagebreak
&&{$f(x,y)= 3 \, x y$}
\\*
\hline
\end{longtable}

\section{LGP and LU classes of bipartite qusext}\label{AP:qusixttClasses}
\noindent
In this appendix we list all the LFP equivalence classes of a bipartite qusext \emph{TEH} states, i.e., states that correspond to polynomial functions. 
We added a \href{https://mfr.de-1.osf.io/render?url=https://osf.io/hn6k8/?direct\%26mode=render\%26action=download\%26mode=render} {comprehensive list of matrices} to an online repository.
\begin{longtable}{ |c|c|l| }
\caption{The full classification of the equivalences of bipartite \emph{TEH} states und LU and LFP operations in dimension $6$. 
The horizontal lines indicate different LFP classes while the singular values identify different LU classes. \label{table:d6classes}}
\\
\hline
{Schmidt Rank} & Singular Values & Polynomial \\
\hline
\endfirsthead
\hline
Schmidt Rank & Singular Values & Polynomial \\
\hline
\endhead
\nopagebreak
\multirow{1}{4em}{$1$}&\multirow{1}{*}{{$\left(1.0,\,0.0,\,0.0,\,0.0,\,0.0,\,0.0\right)$}}&{$f(x,y)= 0$}\Tstrut
\\*
\hline
\nopagebreak
\multirow{1}{4em}{$2$}&\multirow{1}{*}{{$\left(0.70711,\,0.70711,\,0.0,\,0.0,\,0.0,\,0.0\right)$}}&{$f(x,y)= 3 \, x y$}\Tstrut
\\*
\hline
\nopagebreak
\multirow{3}{4em}{$2$}&\multirow{3}{*}{{$\left(0.90506,\,0.42527,\,0.0,\,0.0,\,0.0,\,0.0\right)$}}&{$f(x,y)= 2 \, x^{2} y^{2}$}\Tstrut
\\*
\nopagebreak
&&{$f(x,y)= 2 \, x^{2} y^{2} + 2 \, x y^{2}$}
\\*
\nopagebreak
&&{$f(x,y)= 2 \, x^{2} y^{2} + x y^{2} + 3 \, x y$}
\\*
\hline
\nopagebreak
\multirow{6}{4em}{$2$}&\multirow{6}{*}{{$\left(0.90506,\,0.42527,\,0.0,\,0.0,\,0.0,\,0.0\right)$}}&{$f(x,y)= x^{2} y^{2} + x^{2} y$}\Tstrut
\\*
\nopagebreak
&&{$f(x,y)= x^{2} y^{2} + x^{2} y + x y^{2} + x y$}
\\*
\nopagebreak
&&{$f(x,y)= x^{2} y^{2} + 2 \, x^{2} y + 2 \, x y^{2} + x y$}
\\*
\nopagebreak
&&{$f(x,y)= x^{2} y^{2} + 2 \, x^{2} y + x y^{2} + 2 \, x y$}
\\*
\nopagebreak
&&{$f(x,y)= x^{2} y^{2} + x^{2} y + 2 \, x y^{2} + 2 \, x y$}
\\*
\nopagebreak
&&{$f(x,y)= x^{2} y^{2} + 2 \, x^{2} y + 3 \, x y$}
\\*
\hline
\nopagebreak
\multirow{6}{4em}{$2$}&\multirow{6}{*}{{$\left(0.90506,\,0.42527,\,0.0,\,0.0,\,0.0,\,0.0\right)$}}&{$f(x,y)= 2 \, x^{2} y^{2} + 2 \, x^{2} y$}\Tstrut
\\*
\nopagebreak
&&{$f(x,y)= 2 \, x^{2} y^{2} + 2 \, x^{2} y + x y^{2} + x y$}
\\*
\nopagebreak
&&{$f(x,y)= 2 \, x^{2} y^{2} + x^{2} y + 2 \, x y^{2} + x y$}
\\*
\nopagebreak
&&{$f(x,y)= 2 \, x^{2} y^{2} + x^{2} y + x y^{2} + 2 \, x y$}
\\*
\nopagebreak
&&{$f(x,y)= 2 \, x^{2} y^{2} + 2 \, x^{2} y + 2 \, x y^{2} + 2 \, x y$}
\\*
\nopagebreak
&&{$f(x,y)= 2 \, x^{2} y^{2} + x^{2} y + 3 \, x y$}
\\*
\hline
\nopagebreak
\multirow{3}{4em}{$2$}&\multirow{3}{*}{{$\left(0.90506,\,0.42527,\,0.0,\,0.0,\,0.0,\,0.0\right)$}}&{$f(x,y)= x^{2} y^{2} + x y^{2}$}\Tstrut
\\*
\nopagebreak
&&{$f(x,y)= x^{2} y^{2} + 3 \, x y$}
\\*
\nopagebreak
&&{$f(x,y)= x^{2} y^{2} + 2 \, x y^{2} + 3 \, x y$}
\\*
\hline
\nopagebreak
\multirow{6}{4em}{$2$}&\multirow{6}{*}{{$\left(0.8165,\,0.57735,\,0.0,\,0.0,\,0.0,\,0.0\right)$}}&{$f(x,y)= 2 \, x^{2} y$}\Tstrut
\\*
\nopagebreak
&&{$f(x,y)= x^{2} y + x y$}
\\*
\nopagebreak
&&{$f(x,y)= 2 \, x^{2} y + 2 \, x y$}
\\*
\nopagebreak
&&{$f(x,y)= x^{2} y + 3 \, x y$}
\\*
\nopagebreak
&&{$f(x,y)= 2 \, x^{2} y + 4 \, x y$}
\\*
\nopagebreak
&&{$f(x,y)= x^{2} y + 5 \, x y$}
\\*
\hline
\nopagebreak
\multirow{2}{4em}{$2$}&\multirow{2}{*}{{$\left(0.8165,\,0.57735,\,0.0,\,0.0,\,0.0,\,0.0\right)$}}&{$f(x,y)= 2 \, x y^{2}$}\Tstrut
\\*
\nopagebreak
&&{$f(x,y)= x y^{2} + 3 \, x y$}
\\*
\hline
\nopagebreak
\multirow{4}{4em}{$2$}&\multirow{4}{*}{{$\left(0.8165,\,0.57735,\,0.0,\,0.0,\,0.0,\,0.0\right)$}}&{$f(x,y)= x y^{2} + x y$}\Tstrut
\\*
\nopagebreak
&&{$f(x,y)= 2 \, x y^{2} + 2 \, x y$}
\\*
\nopagebreak
&&{$f(x,y)= 2 \, x y^{2} + 4 \, x y$}
\\*
\nopagebreak
&&{$f(x,y)= x y^{2} + 5 \, x y$}
\\*
\hline
\nopagebreak
\multirow{12}{4em}{$3$}&\multirow{12}{*}{{$\left(0.84403,\,0.4491,\,0.29313,\,0.0,\,0.0,\,0.0\right)$}}&{$f(x,y)= x^{2} y + x y^{2}$}\Tstrut
\\*
\nopagebreak
&&{$f(x,y)= 2 \, x^{2} y + 2 \, x y^{2}$}
\\*
\nopagebreak
&&{$f(x,y)= 2 \, x^{2} y + x y^{2} + x y$}
\\*
\nopagebreak
&&{$f(x,y)= x^{2} y + 2 \, x y^{2} + x y$}
\\*
\nopagebreak
&&{$f(x,y)= x^{2} y + x y^{2} + 2 \, x y$}
\\*
\nopagebreak
&&{$f(x,y)= 2 \, x^{2} y + 2 \, x y^{2} + 2 \, x y$}
\\*
\nopagebreak
&&{$f(x,y)= 2 \, x^{2} y + x y^{2} + 3 \, x y$}
\\*
\nopagebreak
&&{$f(x,y)= x^{2} y + 2 \, x y^{2} + 3 \, x y$}
\\*
\nopagebreak
&&{$f(x,y)= x^{2} y + x y^{2} + 4 \, x y$}
\\*
\nopagebreak
&&{$f(x,y)= 2 \, x^{2} y + 2 \, x y^{2} + 4 \, x y$}
\\*
\nopagebreak
&&{$f(x,y)= 2 \, x^{2} y + x y^{2} + 5 \, x y$}
\\*
\nopagebreak
&&{$f(x,y)= x^{2} y + 2 \, x y^{2} + 5 \, x y$}
\\*
\hline
\pagebreak[4]
\multirow{12}{4em}{$3$}&\multirow{12}{*}{{$\left(0.77814,\,0.57735,\,0.24732,\,0.0,\,0.0,\,0.0\right)$}}&{$f(x,y)= 2 \, x^{2} y^{2} + x^{2} y + x y^{2}$}\Tstrut
\\*
\nopagebreak
&&{$f(x,y)= 2 \, x^{2} y^{2} + 2 \, x^{2} y + 2 \, x y^{2}$}
\\*
\nopagebreak
&&{$f(x,y)= 2 \, x^{2} y^{2} + x^{2} y + x y$}
\\*
\nopagebreak
&&{$f(x,y)= 2 \, x^{2} y^{2} + 2 \, x^{2} y + 2 \, x y$}
\\*
\nopagebreak
&&{$f(x,y)= 2 \, x^{2} y^{2} + 2 \, x^{2} y + x y^{2} + 3 \, x y$}
\\*
\nopagebreak
&&{$f(x,y)= 2 \, x^{2} y^{2} + x^{2} y + 2 \, x y^{2} + 3 \, x y$}
\\*
\nopagebreak
&&{$f(x,y)= 2 \, x^{2} y^{2} + 2 \, x^{2} y + 4 \, x y$}
\\*
\nopagebreak
&&{$f(x,y)= 2 \, x^{2} y^{2} + x^{2} y + x y^{2} + 4 \, x y$}
\\*
\nopagebreak
&&{$f(x,y)= 2 \, x^{2} y^{2} + 2 \, x^{2} y + 2 \, x y^{2} + 4 \, x y$}
\\*
\nopagebreak
&&{$f(x,y)= 2 \, x^{2} y^{2} + x^{2} y + 5 \, x y$}
\\*
\nopagebreak
&&{$f(x,y)= 2 \, x^{2} y^{2} + 2 \, x^{2} y + x y^{2} + 5 \, x y$}
\\*
\nopagebreak
&&{$f(x,y)= 2 \, x^{2} y^{2} + x^{2} y + 2 \, x y^{2} + 5 \, x y$}
\\*
\hline
\nopagebreak
\multirow{12}{4em}{$3$}&\multirow{12}{*}{{$\left(0.77814,\,0.57735,\,0.24732,\,0.0,\,0.0,\,0.0\right)$}}&{$f(x,y)= x^{2} y^{2} + 2 \, x^{2} y + x y^{2}$}\Tstrut
\\*
\nopagebreak
&&{$f(x,y)= x^{2} y^{2} + x^{2} y + 2 \, x y^{2}$}
\\*
\nopagebreak
&&{$f(x,y)= x^{2} y^{2} + 2 \, x^{2} y + x y$}
\\*
\nopagebreak
&&{$f(x,y)= x^{2} y^{2} + x^{2} y + 2 \, x y$}
\\*
\nopagebreak
&&{$f(x,y)= x^{2} y^{2} + x^{2} y + x y^{2} + 3 \, x y$}
\\*
\nopagebreak
&&{$f(x,y)= x^{2} y^{2} + 2 \, x^{2} y + 2 \, x y^{2} + 3 \, x y$}
\\*
\nopagebreak
&&{$f(x,y)= x^{2} y^{2} + x^{2} y + 4 \, x y$}
\\*
\nopagebreak
&&{$f(x,y)= x^{2} y^{2} + 2 \, x^{2} y + x y^{2} + 4 \, x y$}
\\*
\nopagebreak
&&{$f(x,y)= x^{2} y^{2} + x^{2} y + 2 \, x y^{2} + 4 \, x y$}
\\*
\nopagebreak
&&{$f(x,y)= x^{2} y^{2} + 2 \, x^{2} y + 5 \, x y$}
\\*
\nopagebreak
&&{$f(x,y)= x^{2} y^{2} + x^{2} y + x y^{2} + 5 \, x y$}
\\*
\nopagebreak
&&{$f(x,y)= x^{2} y^{2} + 2 \, x^{2} y + 2 \, x y^{2} + 5 \, x y$}
\\*
\hline
\nopagebreak
\multirow{6}{4em}{$3$}&\multirow{6}{*}{{$\left(0.77814,\,0.57735,\,0.24732,\,0.0,\,0.0,\,0.0\right)$}}&{$f(x,y)= x^{2} y^{2} + x y$}\Tstrut
\\*
\nopagebreak
&&{$f(x,y)= x^{2} y^{2} + 2 \, x y^{2} + x y$}
\\*
\nopagebreak
&&{$f(x,y)= x^{2} y^{2} + x y^{2} + 2 \, x y$}
\\*
\nopagebreak
&&{$f(x,y)= x^{2} y^{2} + x y^{2} + 4 \, x y$}
\\*
\nopagebreak
&&{$f(x,y)= x^{2} y^{2} + 5 \, x y$}
\\*
\nopagebreak
&&{$f(x,y)= x^{2} y^{2} + 2 \, x y^{2} + 5 \, x y$}
\\*
\hline
\nopagebreak
\multirow{6}{4em}{$3$}&\multirow{6}{*}{{$\left(0.77814,\,0.57735,\,0.24732,\,0.0,\,0.0,\,0.0\right)$}}&{$f(x,y)= 2 \, x^{2} y^{2} + x y^{2} + x y$}\Tstrut
\\*
\nopagebreak
&&{$f(x,y)= 2 \, x^{2} y^{2} + 2 \, x y$}
\\*
\nopagebreak
&&{$f(x,y)= 2 \, x^{2} y^{2} + 2 \, x y^{2} + 2 \, x y$}
\\*
\nopagebreak
&&{$f(x,y)= 2 \, x^{2} y^{2} + 4 \, x y$}
\\*
\nopagebreak
&&{$f(x,y)= 2 \, x^{2} y^{2} + 2 \, x y^{2} + 4 \, x y$}
\\*
\nopagebreak
&&{$f(x,y)= 2 \, x^{2} y^{2} + x y^{2} + 5 \, x y$}
\\*
\hline
\nopagebreak
\multirow{2}{4em}{$3$}&\multirow{2}{*}{{$\left(0.57735,\,0.57735,\,0.57735,\,0.0,\,0.0,\,0.0\right)$}}&{$f(x,y)= 2 \, x y$}\Tstrut
\\*
\nopagebreak
&&{$f(x,y)= 4 \, x y$}
\\*
\hline
\nopagebreak
\multirow{3}{4em}{$4$}&\multirow{3}{*}{{$\left(0.63998,\,0.63998,\,0.30071,\,0.30071,\,0.0,\,0.0\right)$}}&{$f(x,y)= x^{2} y^{2}$}\Tstrut
\\*
\nopagebreak
&&{$f(x,y)= x^{2} y^{2} + 2 \, x y^{2}$}
\\*
\nopagebreak
&&{$f(x,y)= x^{2} y^{2} + x y^{2} + 3 \, x y$}
\\*
\hline
\nopagebreak
\multirow{6}{4em}{$4$}&\multirow{6}{*}{{$\left(0.63998,\,0.63998,\,0.30071,\,0.30071,\,0.0,\,0.0\right)$}}&{$f(x,y)= 2 \, x^{2} y^{2} + x^{2} y$}\Tstrut
\\*
\nopagebreak
&&{$f(x,y)= 2 \, x^{2} y^{2} + 2 \, x^{2} y + 3 \, x y$}
\\*
\nopagebreak
&&{$f(x,y)= 2 \, x^{2} y^{2} + 2 \, x^{2} y + x y^{2} + 4 \, x y$}
\\*
\nopagebreak
&&{$f(x,y)= 2 \, x^{2} y^{2} + x^{2} y + 2 \, x y^{2} + 4 \, x y$}
\\*
\nopagebreak
&&{$f(x,y)= 2 \, x^{2} y^{2} + x^{2} y + x y^{2} + 5 \, x y$}
\\*
\nopagebreak
&&{$f(x,y)= 2 \, x^{2} y^{2} + 2 \, x^{2} y + 2 \, x y^{2} + 5 \, x y$}
\\*
\hline
\nopagebreak
\multirow{6}{4em}{$4$}&\multirow{6}{*}{{$\left(0.63998,\,0.63998,\,0.30071,\,0.30071,\,0.0,\,0.0\right)$}}&{$f(x,y)= x^{2} y^{2} + 2 \, x^{2} y$}\Tstrut
\\*
\nopagebreak
&&{$f(x,y)= x^{2} y^{2} + x^{2} y + 3 \, x y$}
\\*
\nopagebreak
&&{$f(x,y)= x^{2} y^{2} + x^{2} y + x y^{2} + 4 \, x y$}
\\*
\nopagebreak
&&{$f(x,y)= x^{2} y^{2} + 2 \, x^{2} y + 2 \, x y^{2} + 4 \, x y$}
\\*
\nopagebreak
&&{$f(x,y)= x^{2} y^{2} + 2 \, x^{2} y + x y^{2} + 5 \, x y$}
\\*
\nopagebreak
&&{$f(x,y)= x^{2} y^{2} + x^{2} y + 2 \, x y^{2} + 5 \, x y$}
\\*
\hline
\nopagebreak
\multirow{3}{4em}{$4$}&\multirow{3}{*}{{$\left(0.63998,\,0.63998,\,0.30071,\,0.30071,\,0.0,\,0.0\right)$}}&{$f(x,y)= 2 \, x^{2} y^{2} + x y^{2}$}\Tstrut
\\*
\nopagebreak
&&{$f(x,y)= 2 \, x^{2} y^{2} + 3 \, x y$}
\\*
\nopagebreak
&&{$f(x,y)= 2 \, x^{2} y^{2} + 2 \, x y^{2} + 3 \, x y$}
\\*
\hline
\pagebreak[4]
\multirow{6}{4em}{$4$}&\multirow{6}{*}{{$\left(0.57735,\,0.57735,\,0.40825,\,0.40825,\,0.0,\,0.0\right)$}}&{$f(x,y)= x^{2} y$}\Tstrut
\\*
\nopagebreak
&&{$f(x,y)= 2 \, x^{2} y + x y$}
\\*
\nopagebreak
&&{$f(x,y)= x^{2} y + 2 \, x y$}
\\*
\nopagebreak
&&{$f(x,y)= 2 \, x^{2} y + 3 \, x y$}
\\*
\nopagebreak
&&{$f(x,y)= x^{2} y + 4 \, x y$}
\\*
\nopagebreak
&&{$f(x,y)= 2 \, x^{2} y + 5 \, x y$}
\\*
\hline
\nopagebreak
\multirow{2}{4em}{$4$}&\multirow{2}{*}{{$\left(0.57735,\,0.57735,\,0.40825,\,0.40825,\,0.0,\,0.0\right)$}}&{$f(x,y)= x y^{2}$}\Tstrut
\\*
\nopagebreak
&&{$f(x,y)= 2 \, x y^{2} + 3 \, x y$}
\\*
\hline
\nopagebreak
\multirow{4}{4em}{$4$}&\multirow{4}{*}{{$\left(0.57735,\,0.57735,\,0.40825,\,0.40825,\,0.0,\,0.0\right)$}}&{$f(x,y)= 2 \, x y^{2} + x y$}\Tstrut
\\*
\nopagebreak
&&{$f(x,y)= x y^{2} + 2 \, x y$}
\\*
\nopagebreak
&&{$f(x,y)= x y^{2} + 4 \, x y$}
\\*
\nopagebreak
&&{$f(x,y)= 2 \, x y^{2} + 5 \, x y$}
\\*
\hline
\nopagebreak
\multirow{12}{4em}{$6$}&\multirow{12}{*}{{$\left(0.55023,\,0.55023,\,0.40825,\,0.40825,\,0.17488,\,0.17488\right)$}}&{$f(x,y)= x^{2} y^{2} + x^{2} y + x y^{2}$}\Tstrut
\\*
\nopagebreak
&&{$f(x,y)= x^{2} y^{2} + 2 \, x^{2} y + 2 \, x y^{2}$}
\\*
\nopagebreak
&&{$f(x,y)= x^{2} y^{2} + x^{2} y + x y$}
\\*
\nopagebreak
&&{$f(x,y)= x^{2} y^{2} + 2 \, x^{2} y + x y^{2} + x y$}
\\*
\nopagebreak
&&{$f(x,y)= x^{2} y^{2} + x^{2} y + 2 \, x y^{2} + x y$}
\\*
\nopagebreak
&&{$f(x,y)= x^{2} y^{2} + 2 \, x^{2} y + 2 \, x y$}
\\*
\nopagebreak
&&{$f(x,y)= x^{2} y^{2} + x^{2} y + x y^{2} + 2 \, x y$}
\\*
\nopagebreak
&&{$f(x,y)= x^{2} y^{2} + 2 \, x^{2} y + 2 \, x y^{2} + 2 \, x y$}
\\*
\nopagebreak
&&{$f(x,y)= x^{2} y^{2} + 2 \, x^{2} y + x y^{2} + 3 \, x y$}
\\*
\nopagebreak
&&{$f(x,y)= x^{2} y^{2} + x^{2} y + 2 \, x y^{2} + 3 \, x y$}
\\*
\nopagebreak
&&{$f(x,y)= x^{2} y^{2} + 2 \, x^{2} y + 4 \, x y$}
\\*
\nopagebreak
&&{$f(x,y)= x^{2} y^{2} + x^{2} y + 5 \, x y$}
\\*
\hline
\nopagebreak
\multirow{12}{4em}{$6$}&\multirow{12}{*}{{$\left(0.55023,\,0.55023,\,0.40825,\,0.40825,\,0.17488,\,0.17488\right)$}}&{$f(x,y)= 2 \, x^{2} y^{2} + 2 \, x^{2} y + x y^{2}$}\Tstrut
\\*
\nopagebreak
&&{$f(x,y)= 2 \, x^{2} y^{2} + x^{2} y + 2 \, x y^{2}$}
\\*
\nopagebreak
&&{$f(x,y)= 2 \, x^{2} y^{2} + 2 \, x^{2} y + x y$}
\\*
\nopagebreak
&&{$f(x,y)= 2 \, x^{2} y^{2} + x^{2} y + x y^{2} + x y$}
\\*
\nopagebreak
&&{$f(x,y)= 2 \, x^{2} y^{2} + 2 \, x^{2} y + 2 \, x y^{2} + x y$}
\\*
\nopagebreak
&&{$f(x,y)= 2 \, x^{2} y^{2} + x^{2} y + 2 \, x y$}
\\*
\nopagebreak
&&{$f(x,y)= 2 \, x^{2} y^{2} + 2 \, x^{2} y + x y^{2} + 2 \, x y$}
\\*
\nopagebreak
&&{$f(x,y)= 2 \, x^{2} y^{2} + x^{2} y + 2 \, x y^{2} + 2 \, x y$}
\\*
\nopagebreak
&&{$f(x,y)= 2 \, x^{2} y^{2} + x^{2} y + x y^{2} + 3 \, x y$}
\\*
\nopagebreak
&&{$f(x,y)= 2 \, x^{2} y^{2} + 2 \, x^{2} y + 2 \, x y^{2} + 3 \, x y$}
\\*
\nopagebreak
&&{$f(x,y)= 2 \, x^{2} y^{2} + x^{2} y + 4 \, x y$}
\\*
\nopagebreak
&&{$f(x,y)= 2 \, x^{2} y^{2} + 2 \, x^{2} y + 5 \, x y$}
\\*
\hline
\nopagebreak
\multirow{6}{4em}{$6$}&\multirow{6}{*}{{$\left(0.55023,\,0.55023,\,0.40825,\,0.40825,\,0.17488,\,0.17488\right)$}}&{$f(x,y)= 2 \, x^{2} y^{2} + x y$}\Tstrut
\\*
\nopagebreak
&&{$f(x,y)= 2 \, x^{2} y^{2} + 2 \, x y^{2} + x y$}
\\*
\nopagebreak
&&{$f(x,y)= 2 \, x^{2} y^{2} + x y^{2} + 2 \, x y$}
\\*
\nopagebreak
&&{$f(x,y)= 2 \, x^{2} y^{2} + x y^{2} + 4 \, x y$}
\\*
\nopagebreak
&&{$f(x,y)= 2 \, x^{2} y^{2} + 5 \, x y$}
\\*
\nopagebreak
&&{$f(x,y)= 2 \, x^{2} y^{2} + 2 \, x y^{2} + 5 \, x y$}
\\*
\hline
\nopagebreak
\multirow{6}{4em}{$6$}&\multirow{6}{*}{{$\left(0.55023,\,0.55023,\,0.40825,\,0.40825,\,0.17488,\,0.17488\right)$}}&{$f(x,y)= x^{2} y^{2} + x y^{2} + x y$}\Tstrut
\\*
\nopagebreak
&&{$f(x,y)= x^{2} y^{2} + 2 \, x y$}
\\*
\nopagebreak
&&{$f(x,y)= x^{2} y^{2} + 2 \, x y^{2} + 2 \, x y$}
\\*
\nopagebreak
&&{$f(x,y)= x^{2} y^{2} + 4 \, x y$}
\\*
\nopagebreak
&&{$f(x,y)= x^{2} y^{2} + 2 \, x y^{2} + 4 \, x y$}
\\*
\nopagebreak
&&{$f(x,y)= x^{2} y^{2} + x y^{2} + 5 \, x y$}
\\*
\hline
\pagebreak[4]
\multirow{12}{4em}{$6$}&\multirow{12}{*}{{$\left(0.59682,\,0.59682,\,0.31756,\,0.31756,\,0.20727,\,0.20727\right)$}}&{$f(x,y)= 2 \, x^{2} y + x y^{2}$}\Tstrut
\\*
\nopagebreak
&&{$f(x,y)= x^{2} y + 2 \, x y^{2}$}
\\*
\nopagebreak
&&{$f(x,y)= x^{2} y + x y^{2} + x y$}
\\*
\nopagebreak
&&{$f(x,y)= 2 \, x^{2} y + 2 \, x y^{2} + x y$}
\\*
\nopagebreak
&&{$f(x,y)= 2 \, x^{2} y + x y^{2} + 2 \, x y$}
\\*
\nopagebreak
&&{$f(x,y)= x^{2} y + 2 \, x y^{2} + 2 \, x y$}
\\*
\nopagebreak
&&{$f(x,y)= x^{2} y + x y^{2} + 3 \, x y$}
\\*
\nopagebreak
&&{$f(x,y)= 2 \, x^{2} y + 2 \, x y^{2} + 3 \, x y$}
\\*
\nopagebreak
&&{$f(x,y)= 2 \, x^{2} y + x y^{2} + 4 \, x y$}
\\*
\nopagebreak
&&{$f(x,y)= x^{2} y + 2 \, x y^{2} + 4 \, x y$}
\\*
\nopagebreak
&&{$f(x,y)= x^{2} y + x y^{2} + 5 \, x y$}
\\*
\nopagebreak
&&{$f(x,y)= 2 \, x^{2} y + 2 \, x y^{2} + 5 \, x y$}
\\*
\hline
\nopagebreak
\multirow{2}{4em}{$6$}&\multirow{2}{*}{{$\left(0.40825,\,0.40825,\,0.40825,\,0.40825,\,0.40825,\,0.40825\right)$}}&{$f(x,y)= x y$}\Tstrut
\\*
\nopagebreak
&&{$f(x,y)= 5 \, x y$}
\\*
\hline
\end{longtable}

\end{document}